\documentclass[lettersize,journal]{IEEEtran}

\usepackage{amsmath,amsfonts,amssymb}
\usepackage{algorithm}
\usepackage{algorithmic}
\usepackage{array}
\usepackage[caption=false,font=normalsize,labelfont=sf,textfont=sf]{subfig}
\usepackage{textcomp}
\usepackage{stfloats}
\usepackage{url}
\usepackage{verbatim}
\usepackage{graphicx}
\usepackage{balance}
\usepackage{booktabs}
\usepackage{multirow}
\usepackage{enumitem}
\usepackage[table]{xcolor}
\usepackage{array}
\usepackage{hyperref}
\usepackage{orcidlink}
\usepackage{tcolorbox}
\def\BibTeX{{\rm B\kern-.05em{\sc i\kern-.025em b}\kern-.08em
    T\kern-.1667em\lower.7ex\hbox{E}\kern-.125emX}}

\begin{document}

\title{How Much Feedback Is Needed to Break Template Protection?
The Geometric Vulnerability of Biometric Authentication}
\author{Eliron Rahimi~\orcidlink{0009-0007-4424-2123}, Margarita Osadchy~\orcidlink{0000-0001-5480-5099}, Orr Dunkelman~\orcidlink{0000-0001-5799-2635}
\thanks{Eliron Rahimi is with the Department of Computer Science,
Technion--Israel Institute of Technology, Haifa, Israel
.}
\thanks{Margarita Osadchy and Orr Dunkelman are with the Department of
Computer Science, University of Haifa, Haifa, Israel.}%

\thanks{Corresponding author: Eliron Rahimi, e-mail: elironrahimi@campus.technion.ac.il}
}

\markboth{Preprint}{}

\maketitle

\begin{abstract}

Biometric authentication systems increasingly use privacy-preserving mechanisms to protect sensitive templates. However, the authentication interface itself may still leak information that enables reconstruction attacks. Prior attacks exploit released similarity scores, whereas more advanced systems reveal only the binary accept/reject decision. This binary feedback provides substantially less information, making template reconstruction far more challenging.

In this work, we introduce query-efficient reconstruction attacks under extremely limited feedback, addressing two key limitations of prior attacks. First, we introduce the first attack that reconstructs protected biometric templates using only authentication success/failure feedback by exploiting the geometric constraints of biometric authentication. Second, we introduce distribution-aware query-efficient reconstruction methods for score-releasing systems, achieving a more realistic threat model by reducing the number of required queries by a factor of $5\times$ compared to recent prior score-based attacks. Finally, we combine these two insights by incorporating a pretrained facial diffusion prior into our geometric binary attack, reducing the number of required queries by a factor of $5\times$.

Our findings reveal a fundamental security limitation of biometric authentication: independently of the template protection mechanism, the authentication process itself exposes geometric information that can enable perfect template reconstruction up to numerical precision. Moreover, the distribution of biometric templates can be leveraged to make significantly more effective reconstruction attacks.

Code is available at \href{https://github.com/ElironRahimi/geometric_biometric_reconstruction} {\includegraphics[height=1em]{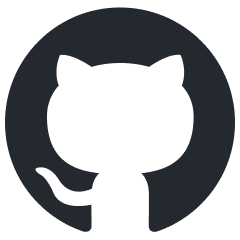}}.

\end{abstract}


\section{Introduction}
\IEEEPARstart{B}{iometric} authentication systems have become a cornerstone of modern identity verification due to their convenience and strong discriminative power. Deep neural network-based feature extractors such as ArcFace~\cite{Deng_2019_CVPR} and FaceNet~\cite{Schroff_2015_CVPR} produce compact and highly descriptive embeddings that enable accurate recognition across a wide range of applications. However, these embeddings inherently encode sensitive biometric information, raising significant privacy and security concerns.

It has been shown that face images can be reconstructed from templates generated by deep feature extraction algorithms~\cite{6199756,8338413}, highlighting the vulnerability of biometric templates. The European Union's General Data Protection Regulation (GDPR)~\cite{regulation2016regulation} declared biometric data as sensitive personal data, and ISO/IEC 24745~\cite{rane2014standardization} provided a standardized framework for protecting biometric information through the requirements of unlinkability, irreversibility, and renewability. In this work, we address the irreversibility requirement, namely, the requirement that the protected representation should not allow recovery of the original biometric template.

To address these risks, a large body of work has focused on protecting biometric templates using cryptographic techniques such as Fully Homomorphic Encryption (FHE) and related privacy-preserving protocols~\cite{9980432,8698601,8897272,9794523,9724306}. FHE is particularly attractive because it enables matching directly in the encrypted domain without decrypting the stored template, and can preserve recognition accuracy while reducing direct exposure of sensitive data~\cite{9585508,10347446,9211046,9548305}.

A common assumption underlying these systems is that limiting the information revealed during authentication enhances security. For example, the first-generation FHE-based biometric systems returned similarity scores computed in the encrypted domain, typically squared Euclidean distance or cosine similarity, between the stored template and the query template~\cite{9794523,8698601,8897272,9678986,9548305}. More recent systems improved privacy by performing threshold comparison within the protected domain and returning only a binary authentication result, i.e., accept or reject~\cite{9585508,10347446,9724306}. Intuitively, this binary-output setting appears substantially safer because each query reveals only one bit of information.

Despite these protections, biometric templates may still be vulnerable to reconstruction attacks when some form of feedback is available. When similarity scores are exposed, an adversary can exploit these scores by hill-climbing~\cite{6994220} or by solving systems of equations derived from the observed scores~\cite{4359297,10449211, 11410617, rahimi2025testing}. The security of systems that reveal only binary authentication outcomes is more subtle. The main question is whether a system that reveals the minimal possible authentication feedback (success or failure) still leaks enough information to reconstruct the protected template.

In this paper, we answer this question affirmatively and introduce a geometric approach for reconstructing protected biometric templates from extremely limited feedback. We show that even binary authentication outcomes, which reveal only whether an authentication attempt succeeds or fails, provide sufficient information for accurate template reconstruction. The key observation is that binary decisions define geometric constraints in the feature space: in squared-Euclidean-distance systems, the acceptance region forms a sphere around the target template, while in cosine-similarity systems, it forms a cone on the unit hypersphere. By locating points on the decision boundary, the attacker can construct equations that reveal the protected template.

Beyond demonstrating reconstruction from binary feedback, we study the query efficiency of reconstruction attacks. We therefore develop query-efficient attacks for both score-releasing and binary-feedback systems. For score-releasing systems, we leverage distribution-aware input selection to reduce the number of required authentication queries. For binary-feedback systems, we combine our geometric approach with a generative prior to substantially reduce the number of queries required for accurate reconstruction.

Extensive experiments on state-of-the-art biometric feature extractors show that our methods achieve the lowest reconstruction error while requiring substantially fewer queries than existing approaches across all evaluated benchmarks. These results highlight a fundamental limitation of current template protection mechanisms: the authentication functionality itself leaks sufficient information for reconstruction of the template. Overall, our findings demonstrate that, independent of the underlying template protection mechanism (e.g., homomorphic encryption), securing biometric templates is difficult when repeated authentication queries are possible.

We summarize our main contributions as follows:
\begin{itemize}[leftmargin=15pt]
    \item We introduce the first template reconstruction attack from binary authentication feedback, showing that protected biometric templates can be reconstructed using only accept/reject decisions independent of the template protection mechanism.

    \item We develop prior-based query-efficient reconstruction methods for score-releasing systems, reducing the number of required queries by up to a factor of $5\times$ compared to recent score-based attacks.

    \item We develop a query-efficient binary-feedback attack by integrating our geometric reconstruction approach with a pretrained facial diffusion-based prior, reducing the number of required queries by over $80\%$ in our experiments.

    \item Our findings reveal a fundamental limitation of cryptographic template protection, the authentication process itself exposes geometric information that can enable perfect template reconstruction up to numerical precision, and the distribution of biometric templates can be leveraged to make significantly more effective reconstruction attacks.

\end{itemize}

\section{Related Work}
\label{sec:related_work}

\subsection{Biometric Template Protection}

While biometric systems offer convenience and high
accuracy, they inherently raise serious privacy and security concerns, as biometric identifiers
are unique and irrevocable. Once compromised, they cannot be reissued like passwords or
tokens. To mitigate these risks, the concept of protected biometric systems (also referred to as
biometric template protection schemes) has been introduced.

Protected biometric systems aim to secure stored biometric templates and ensure that no
sensitive information about the original biometric data can be inferred from them. Such systems
are designed around three main requirements: \emph{unlinkability}, \emph{irreversibility}, and \emph{renewability}~\cite{rane2014standardization}.
Unlinkability ensures that protected templates generated from the same biometric data in different systems cannot be cross-matched to reveal user identity. 
Irreversibility guarantees that it is computationally infeasible to reconstruct the original biometric feature vector from its protected representation.
Renewability allows the generation of a new protected template from the same biometric input if a previous one is compromised, analogous to resetting a password.

Template protection approaches are typically categorized into two main paradigms: \emph{feature transformation} and \emph{biometric cryptosystems}.
Feature transformation methods apply a reversible transformation on the feature space using user-specific keys or random projections.
Examples include salting, BioHashing, and cancelable biometrics~\cite{7192838}.
Biometric cryptosystems, on the other hand, bind biometric data to cryptographic keys to derive secure representations, as in fuzzy commitment and fuzzy vault schemes~\cite{9980432,9599594}. More recent approaches employ homomorphic encryption~\cite{8698601,8897272,9724306} or secure multiparty computation~\cite{9980432} to enable privacy-preserving matching without revealing the underlying biometric data.

Despite significant progress, achieving an ideal balance among recognition accuracy, security, and computational efficiency remains challenging.
Protected biometric systems must ensure high recognition performance under intra-user variations while maintaining strong resistance against inversion, cross-matching, and linkage attacks.
Consequently, designing robust and efficient template protection mechanisms continues to be an active area of research in biometric security.
We further detail FHE–based systems, which are considered the most advanced and promising among these solutions. Nevertheless, our results are applicable to any protection mechanism  that preserves the recognition accuracy (e.g., homomorphic encryption).

\subsection{Reconstruction Attacks from Similarity Scores}

Reconstruction attacks exploiting similarity scores have been widely studied. Iterative template adaptation scheme, popularly known as a hill-climbing attack, to break into a biometric system based on match scores. Hill-climbing attacks iteratively submit synthetic representations of the attacked user's biometrics and modify them based on observed scores to improve the matching result~\cite{6994220}.

In contrast to adaptive hill-climbing, Mohanty et al.~\cite{4359297} proposed a non-adaptive method that exploits exposed similarity scores to construct equations from the revealed data. The template is recovered by solving the resulting system of equations. Although this method was developed before deep learning became dominant in biometrics, its basic idea remains applicable to modern feature representations and to protected systems that release scores.

More recent work adapted these ideas to deep feature representations. Bassit et al.~\cite{10449211} proposed a non-adaptive attack for recovering raw templates from leaked comparison scores in homomorphic-encryption-based systems. Their work focuses on cosine similarity scores and uses a random set of fake templates with their matching scores to solve a constrained optimization problem under a unit-norm prior. While effective, such methods typically require a large number of queries and do not explicitly optimize the selection of queries.

Shahreza and Marcel~\cite{10581959} proposed a method for reconstructing face images from protected facial templates, including templates protected by homomorphic encryption. Their threat model assumes complete knowledge of the protection scheme and its secrets, including the decryption key. In contrast, our setting assumes that protected templates cannot be decrypted and that the attacker interacts only with the authentication interface.






\section{Template Reconstruction Attack Methodology}
\label{sec:limited_feedback_attacks}
After defining the adversarial model in Section~\ref{sec:threat_model}, we present a method for recovering biometric templates from similarity scores (Section~\ref{sec:score_reconstruction}). While variants of score-based attacks have been studied in prior work~\cite{10449211}, our version offers improvements in certain settings and employs a unified algebraic framework that covers the two dominant similarity measures in biometric authentication: squared Euclidean distance and cosine similarity. Importantly, a key component of this method is later adapted in our binary-feedback attack, which reconstructs biometric templates using only binary (yes/no) matching outcomes for both geometries induced by squared Euclidean distance and cosine similarity (Section~\ref{sec:binary_reconstruction}).

\subsection{Threat Model}
\label{sec:threat_model}

We consider a threat model in which the feature extractor is run by a semi-honest service provider assumed to follow the protocol correctly but potentially curious about inputs and outputs. The service provider holds the original templates protected under the matching module's or sensor's public key, and cannot decrypt those templates. The attacker has black-box access to the matching module and can submit arbitrary feature vectors. We define the term \emph{breaking-set} as the collection of feature vectors submitted by the attacker to the system, which may be derived from actual biometric samples or crafted independently.

This is a common adversarial model used in prior work~\cite{8698601,8897272,9678986,10449211}, which assumes that the adversary controls the sensor or has access to the feature extractor model in it. Depending on the attack scenario, the attacker receives either similarity scores or binary match/no-match outcomes from the system.

In addition, the attacks rely on the following assumptions:
\begin{itemize}[leftmargin=15pt]
    \item The target template is stored protected on the server (using any protection mechanism) and cannot be accessed by the attacker.
    \item The attacker has access to the feature extractor in order to extract templates from candidate images.
\end{itemize}

\subsection{Algebraic Reconstruction from Similarity Scores}
\label{sec:score_reconstruction}
Let $f = [f_1, f_2, \ldots, f_d] \in \mathbb{R}^d$ represent the target user's biometric template in the plaintext domain created during enrollment and subsequently stored in protected form in the database, where $d$ is the feature-vector dimension. The goal is to find an approximation to $f$. Let $\{q^i\}_{i=1}^{d+1}$ be a breaking-set of $d+1$ independent feature vectors, where $q^i=[q^i_1,q^i_2,\ldots,q^i_d]\in\mathbb{R}^d$.

\noindent\textit{Squared Euclidean distance case:} When the target system's similarity score is the squared Euclidean distance, the attack follows the triangulation principle of~\cite{4359297}. For each $q^i$, the attacker submits an authentication trial with the claimed identity being the target user's ID, totaling $d+1$ independent trials. Let $\{s_i\}_{i=1}^{d+1}$ represent the squared Euclidean distances between the target template $f$ and the $i$-th template in the breaking-set $q^i$, exposed during the authentication trials. Using this data, the attacker constructs the following nonlinear system of equations:
\begin{equation}
\label{eq:system}
\small
(f_1 - q_{1}^i)^2 +  \ldots + (f_d - q_{d}^i)^2 = s_i, \quad i = 1, \dots, d+1.
\end{equation}
This system can be rewritten as
\begin{equation}
\label{eq:system2}
\|f\|^2 - \ell_i = 0, \quad i = 1, \dots, d+1,
\end{equation}
where $\|f\|^2=\sum_{j=1}^d (f_j)^2$ and
\begin{equation}
\ell_i = s_i - \sum_{j=1}^d \left(-2f_j q_j^i + (q_j^i)^2\right)
\end{equation}
is the linear part of the $i$-th equation.

Subtracting one of these equations, e.g., the last one, from all the others eliminates the common nonlinear term and creates a linear system of $d$ equations and $d$ variables:
\begin{equation}
\label{eq:system3}
\ell_{d+1} - \ell_i = 0, \quad i = 1, \dots, d.
\end{equation}
An approximation for $f$ can then be obtained by solving this linear system.

\noindent\textit{Cosine similarity case:} In this case, every authentication trial automatically creates a linear equation derived from the cosine similarity definition and the fact that modern deep feature vectors are normalized:
\begin{equation}
\small
\label{eq:cosine_similarity_def}
\text{cosine similarity}(A,B)=\frac{A\cdot B}{\|A\|\|B\|}.
\end{equation}
Since $\|f\|=\|q^i\|=1$, the $i$-th query yields
\begin{equation}
\label{eq:cosine_similarity}
\small
c_i = \sum_{j=1}^{d} f_j q_j^i,
\end{equation}
where $c_i$ is the cosine similarity score between $f$ and $q^i$. Thus, only $d$ independent authentication trials are needed to construct a linear system of $d$ equations in $d$ variables.

\subsection{Reconstruction from Binary Authentication Feedback}
\label{sec:binary_reconstruction}

In this section, we present our reconstruction attack from binary authentication feedback. Since this is the first reconstruction attack based only on binary authentication responses, we also introduce a simple baseline in Section~\ref{sec:binary_baseline}, which is used for comparison in our evaluation. In Section~\ref{sec:binary_sed} we present the attack for squared-Euclidean-distance systems , where the acceptance region forms a sphere around the target template and reconstruction reduces to recovering the sphere center. We then extend the attack to cosine-similarity systems in Section~\ref{sec:binary_cosine}, where the acceptance region forms a cone on the unit hypersphere and reconstruction reduces to recovering the cone axis.

\subsubsection{Baseline Approach}
\label{sec:binary_baseline}
As a simple baseline, the attacker performs authentication attempts using a breaking-set $B$ containing facial templates from different identities, and averages the feature vectors that are falsely accepted by the system. Formally, the attacker collects all templates $b\in B$ such that $A(b)=1$ and returns their average as the reconstructed template. Geometrically, this method averages points inside the acceptance sphere in SED-based systems, or inside the acceptance cone/spherical cap in cosine-similarity systems. The full procedure is summarized in Algorithm~\ref{alg:naive}.

\begin{algorithm}
\caption{Baseline Approach}
\label{alg:naive}
\begin{algorithmic}[1]
    \REQUIRE Authentication function $A$ and breaking-set $B$
    \STATE $S \gets \{\}$
    \FOR{$b \in B$}
        \IF{$A(b)=1$}
            \STATE $S \gets S \cup \{b\}$
        \ENDIF
    \ENDFOR
    \STATE \textbf{return} the average of all templates in $S$
\end{algorithmic}
\end{algorithm}

\subsubsection{Squared Euclidean Distance Systems}
\label{sec:binary_sed}

\begin{figure*}[ht]
    \centering
    \includegraphics[width=0.9\textwidth]{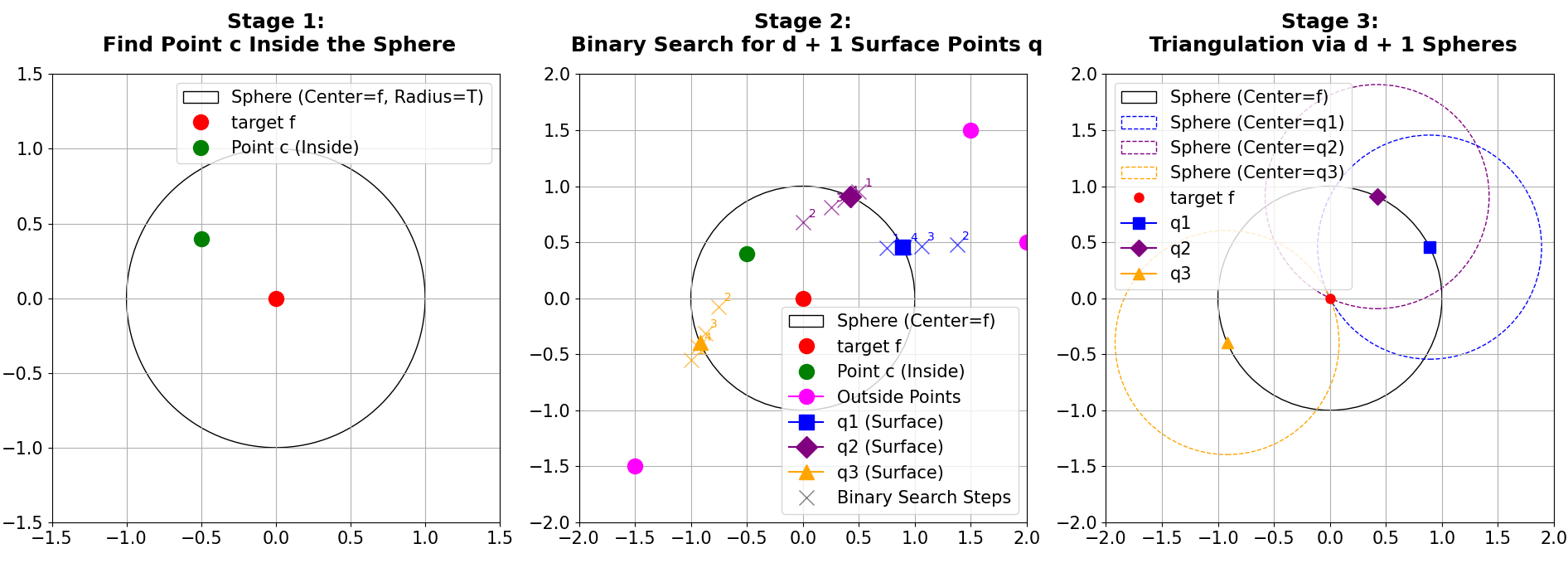}
    \captionsetup{justification=centering}
    \caption{Illustration of our template reconstruction attack using only binary authentication results in 2D.}
    \label{fig:yes_no_attack}
\end{figure*}

Let $T$ denote the system's matching threshold. In SED systems, recovering the target template using yes/no answers is equivalent to the following geometric problem: finding the center of a $d$-dimensional sphere with radius $T$, given access to a function $A:\mathbb{R}^d\to\{0,1\}$ that maps an input $x$ to either 0 or 1:
\begin{equation}
\label{auth_functions}
A(x)=
\begin{cases}
1, & \text{if } x \text{ is inside the sphere},\\
0, & \text{otherwise}.
\end{cases}
\end{equation}

The first stage of the attack is to find one feature vector $c$ that passes authentication, i.e., $A(c)=1$. On average, this stage requires $1/\mathrm{FMR}$ authentication attempts, where FMR is the false match rate.

The second stage finds $d+1$ points on the surface of the sphere. Starting from the valid point $c$, the attacker constructs a point $o_i$ outside the sphere by sampling a random direction and moving a distance $2T$ in that direction. The attacker then performs binary search between the inside point and the outside point using only the binary outputs of the authentication system. As the number of binary-search iterations $P$ increases, the resulting point $q_i$ approaches the surface of the sphere, meaning that its distance from $f$ converges to $T$.

Repeating this process $d+1$ times allows the attacker to construct the nonlinear system in Eq.~\eqref{eq:system}, where all distances $s_1,s_2,\ldots,s_{d+1}$ are equal to $T$. The final stage solves the system as in Section~\ref{sec:score_reconstruction}. 


\begin{algorithm}[t]
\caption{Reconstruction from binary authentication feedback in SED systems}
\label{alg:yes_no}
\begin{algorithmic}[1]
    \REQUIRE Precision parameter $P$, authentication function $A$ as defined in Eq.~\eqref{auth_functions}, threshold $T$, breaking-set $B$.
    \FOR{$b$ in $B$}
        \IF{$A(b)$}
            \STATE $c=b$; break
        \ENDIF
    \ENDFOR
    \STATE Initialize an empty system of equations $S$
    \FOR{$d+1$ iterations}
        \STATE Sample a random vector $r$ with $\|r\|=1$
        \STATE $o_i=c+2Tr$
        \STATE $c_i=c$
        \FOR{$P$ iterations}
            \STATE $q_i=(c_i+o_i)/2$
            \IF{$A(q_i)$}
                \STATE $c_i=q_i$
            \ELSE
                \STATE $o_i=q_i$
            \ENDIF
        \ENDFOR
        \STATE Add $(c_i+o_i)/2$ to $S$ as a point at distance $T$ from the center
    \ENDFOR
    \STATE Solve $S$ for the center of the sphere
    \STATE \textbf{return} the solution of $S$
\end{algorithmic}
\end{algorithm}

The total number of authentication attempts needed for reconstruction with negligible reconstruction loss is
\begin{equation}
\label{eq:sed_binary_query_complexity}
\frac{1}{\mathrm{FMR}} + P(d+1).
\end{equation}

\subsubsection{Cosine Similarity Systems}
\label{sec:binary_cosine}
\begin{figure*}[ht]
    \centering
    \includegraphics[width=1\textwidth]{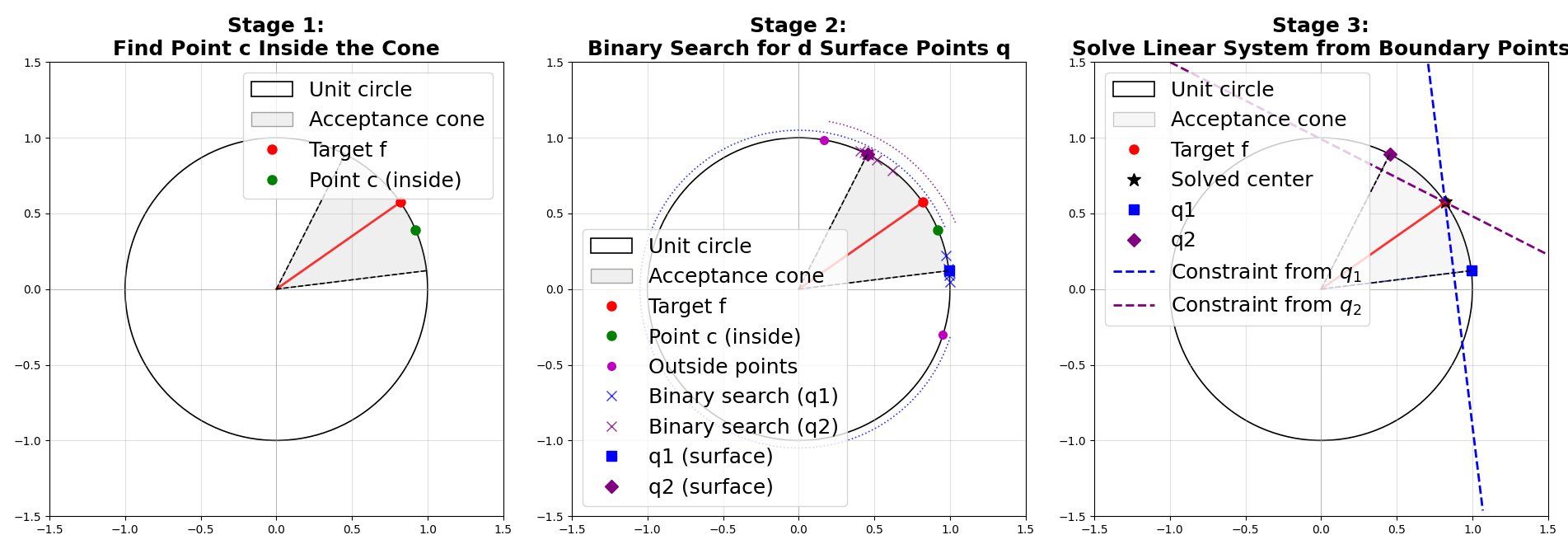}
    \captionsetup{justification=centering}
    \caption{Illustration of the reconstruction attack from binary authentication feedback in cosine similarity systems in 2D.}
    \label{fig:yes_no_cosine_attack}
\end{figure*}
We extend the binary-feedback reconstruction attack to biometric systems that use cosine similarity. In such systems, authentication is determined according to whether the cosine similarity with the target template exceeds a threshold. Geometrically, the accepted set is no longer a sphere around the target template, but rather a cone intersected with the unit hypersphere. Therefore, the reconstruction problem under binary authentication feedback can be viewed as recovering the axis of this cone using only yes/no authentication outcomes.

Similarly to the SED case, we assume that the attacker can query the system through a function $A:\mathbb{R}^d\to\{0,1\}$, where
\begin{equation}
\label{eq:cosine_auth_function}
A(x)=
\begin{cases}
1, & \text{if } \cos(x,f)\geq T,\\
0, & \text{otherwise},
\end{cases}
\end{equation}
where $f\in\mathbb{R}^d$ is the unknown target template and $T$ is the system threshold. Since all feature vectors are normalized, the acceptance region corresponds to a spherical cap on the unit hypersphere, or equivalently to the interior of a cone whose axis is $f$.

Our method follows the same logic as the SED case, with two key modifications. First, instead of stepping outside a sphere, we move outside the acceptance cone while remaining on the unit hypersphere by taking a random angular step from a valid point $c$, producing a point $o_i$ outside the cone. Second, since cosine similarity operates on normalized vectors, all midpoints computed during binary search are re-normalized before evaluation.

As in the previous case, finding such a point $c$ requires on average $1/\mathrm{FMR}$ queries. For each of the $d$ equations, we generate a point $o_i$ outside the cone and perform binary search between $c_i$ and $o_i$, normalizing each midpoint. As $P$ increases, the resulting point $q_i$ converges to the boundary, yielding a linear constraint
\begin{equation}
\label{eq:cosine_boundary_constraint}
q_i^\top f = T.
\end{equation}
Repeating this process $d$ times produces a linear system that can be solved to recover $f$. Unlike the SED case, no equation subtraction is required, as the formulation is already linear.

\begin{algorithm}[t]
\caption{Reconstruction from binary authentication feedback in cosine similarity systems}
\label{alg:yes_no_cosine}
\begin{algorithmic}[1]
    \REQUIRE Precision parameter $P$, authentication function $A$ as defined in Eq.~\eqref{eq:cosine_auth_function}, threshold $T$, breaking-set $B$
    \FOR{$b$ in $B$}
        \IF{$A(b)$}
            \STATE $c=b$; break
        \ENDIF
    \ENDFOR
    \STATE Initialize an empty system of equations $S$
    \FOR{$d$ iterations}
        \STATE Sample a random unit vector $u$ such that $u^\top c=0$
        \STATE Generate a point outside the cone using $o_i=\cos(\theta)c+\sin(\theta)u$
        \STATE $c_i=c$
        \FOR{$P$ iterations}
            \STATE $q_i=(c_i+o_i)/2$
            \STATE $q_i=q_i/\|q_i\|$
            \IF{$A(q_i)$}
                \STATE $c_i=q_i$
            \ELSE
                \STATE $o_i=q_i$
            \ENDIF
        \ENDFOR
        \STATE Add the equation $q_i^\top f=T$ to $S$
    \ENDFOR
    \STATE Solve $S$ for $f$
    \STATE \textbf{return} the solution of $S$
\end{algorithmic}
\end{algorithm}

To generate a point outside the acceptance cone, we rotate an accepted,
normalized feature vector $x$ along a random unit direction $u$ orthogonal
to $x$, such that $u^\top x=0$. The resulting point is
\begin{equation}
\label{eq:random_step_cosine}
n = \cos(\theta)x + \sin(\theta)u,
\end{equation}
where $\theta=2\arccos(T)+\varepsilon$, with
$0<\varepsilon<\pi-2\arccos(T)$ and $T>0$.
Since any accepted point lies within $\arccos(T)$ of the target,
this choice guarantees that $n$ lies outside the acceptance cone,
regardless of the sampled direction, while preserving its unit norm.

The total number of authentication attempts required by this attack is, on average,
\begin{equation}
\label{eq:cosine_binary_query_complexity}
\frac{1}{\mathrm{FMR}} + Pd,
\end{equation}
which has the same asymptotic complexity as score-based cosine reconstruction while requiring only binary authentication feedback.

\section{Query-Efficient Template Reconstruction Attack Methodology}
\label{sec:query_efficient}

\subsection{Query-Constrained Threat Model}

We adopt the threat model as defined in Section~\ref{sec:threat_model}, where the attacker has black-box access to the system and receives limited feedback, either similarity scores or binary authentication outcomes. In addition, we impose a query budget constraint, restricting the attacker to at most $Q$ authentication attempts.

We define the breaking-set $B$ as the set of feature vectors submitted by the attacker to the system, with $|B|\leq Q$. Thus, each query corresponds to adding one element to $B$. Under this constraint, the attacker's objective is to design a breaking-set and an attack strategy that minimize the reconstruction error of the target template given the budget $Q$. This formulation highlights the trade-off between reconstruction accuracy and query efficiency.

\subsection{Query-Efficient Reconstruction from Similarity Scores}
\label{sec:query_efficient_scores}

The score-based reconstruction methods in Section~\ref{sec:score_reconstruction} become exact when $d+1$ independent equations are available. However, under a strict query budget, the attacker may have fewer equations than the dimension of the template. In this setting, reconstruction quality strongly depends  on the attack strategy.

Bassit et al.~\cite{10449211} use a random set of fake templates and their matching scores with the target template to perform recovery. They leverage the prior knowledge that the target template has unit norm, as deep feature extractors produce normalized embeddings, and formulate the constrained optimization problem as
\begin{equation}
\hat{x} = \arg\min_x \|Ax-b\|^2 \quad \text{subject to } \|x\|_2=1.
\end{equation}

We propose a stronger prior that constrains the recovered template to the biometric-template distribution.Formally, our objective becomes
\begin{equation}
\label{eq:dist_equation}
\hat{x} = \arg\min_{x\sim\mathcal{D}} \|Ax-b\|^2,
\end{equation}

where $\mathcal{D}$ denotes the distribution of the biometric templates. This prior is especially important in the query-limited setting, where the number of equations may be smaller than the number of unknowns (the template dimension). In such underdetermined systems, many vectors can satisfy the observed constraints, but only a small subset falls within the biometric-template distribution. Restricting the solution to this distribution therefore narrows the feasible space and leads to more accurate reconstruction. 

To approximate the objective in Eq.~\eqref{eq:dist_equation}, we develop an input-selection algorithm based on a simple distributional observation: Let $X=\{x_i\}_{i=1}^{n}$ be a set of facial templates extracted from real images using the same feature extractor as the target template. A breaking-set selected from $X$ is more informative than one constructed by sampling random vectors in $\mathbb{R}^d$ (as used in ~\cite{10449211}), since samples from $X$ follow the feature extractor's template distribution, they better span the region in which the target template is expected to reside. This provides a built-in prior, as the linear structure of the feature space encourages consistent geometric relationships between samples.

We further use $X$ to construct a breaking-set that spans the target template even more effectively than simply selecting templates from $X$. We aim to avoid redundant queries by ensuring that the selected vectors span diverse directions in the feature space. Since the feature vectors are $\ell_2$-normalized, cosine similarity is equivalent to the dot product, and PCA becomes a natural tool for spanning-set construction.

Formally, let $X\in\mathbb{R}^{n\times d}$ denote a dataset of
normalized biometric templates, where $n$ is the number of available
samples and $d$ is the embedding dimension. Let
$\mu=\frac{1}{n}\sum_{i=1}^{n}x_i$ denote the empirical mean and let
$X_c$ denote the mean-centered template matrix. PCA computes the
eigen-decomposition of the empirical covariance matrix
\begin{equation}
C = \frac{1}{n}X_c^\top X_c.
\end{equation}
The eigenvectors of $C$, denoted by $\{v_1,v_2,\ldots,v_d\}$, form an orthonormal basis of directions ordered by explained variance. Selecting the first $k$ eigenvectors gives the $k$ most informative and diverse directions in which the dataset varies. Given a query budget $B$, we use the first $B$ principal components as the queries. When $B<d$, this yields a low-rank approximation of the target template that minimizes reconstruction error in terms of squared Euclidean distance, which coincides with cosine-based reconstruction error due to normalization. To prevent wasted queries, we use a variance threshold $\tau>0$. If the eigenvalue $\lambda_i$ associated with eigenvector $v_i$ satisfies $\lambda_i<\tau$, this means that the direction provides negligible information. In such cases, instead of using $v_i$, we replace it with a randomly sampled template from the dataset $X$. Algorithm~\ref{alg:pca-selection} summarizes the full selection process. 

After constructing the breaking-set, the resulting constraints can be solved using a least-squares linear solver, providing an efficient approximation to the objective in Eq.~\eqref{eq:dist_equation} only by constructing the breaking-set.

\begin{algorithm}[t]
\caption{PCA-based input selection}
\label{alg:pca-selection}
\begin{algorithmic}[1]
    \REQUIRE Dataset $X\in\mathbb{R}^{n\times d}$ of normalized templates, query budget $B$, variance threshold $\tau$
    \STATE Compute the empirical mean $\mu=\frac{1}{n}\sum_{i=1}^{n}x_i$
    \STATE Center the dataset: $X_c=X-\mathbf{1}\mu^\top$
    \STATE Compute the covariance matrix $C=\frac{1}{n-1}X_c^\top X_c$
    \STATE Compute the eigen-decomposition $C=V\Lambda V^\top$
    \STATE Sort the eigenvectors $\{v_i\}$ by eigenvalues $\{\lambda_i\}$ in descending order
    \STATE Initialize breaking-set $S=\{\}$
    \FOR{$i=1$ to $B$}
        \IF{$\lambda_i\geq\tau$}
            \STATE Add the normalized $v_i$ to $S$
        \ELSE
            \STATE Select a random template $x\sim X$ and add it to $S$
        \ENDIF
    \ENDFOR
    \STATE \textbf{return} $S$
\end{algorithmic}
\end{algorithm}

\subsection{Query-Efficient Reconstruction from Binary Authentication Feedback}
\label{sec:query_efficient_binary}

As shown in Section~\ref{sec:binary_reconstruction}, the number of queries required for near-perfect reconstruction using binary authentication results is $1/\mathrm{FMR}+P(d+1)$ in the SED case and $1/\mathrm{FMR}+Pd$ in the cosine case. The term $1/\mathrm{FMR}$ is determined by the system threshold, $P$ controls the binary-search precision, and $d$ controls the number of equations required by the algebraic solution. Our goal is to reduce the number of required equations while preserving accurate reconstruction. This directly reduces the query complexity: each removed equation eliminates one boundary-search procedure, saving $P$ authentication queries. We therefore aim to replace the purely algebraic solver with a prior-based reconstruction method that can recover accurate templates from fewer equations.

The input-selection strategy presented in Section~\ref{sec:query_efficient_scores} improves reconstruction when the attacker can choose informative score queries. In the binary setting, however, the equations are generated through a random boundary-search process, and the attacker cannot directly choose the final boundary points to align with the most informative feature-space directions. Therefore, input selection alone cannot be applied in the same manner.

To overcome this limitation, we approximate the objective in Eq.~\eqref{eq:dist_equation} by formulating reconstruction as an optimization problem that incorporates prior information. Specifically, we seek a template $\hat{x}$ that minimizes
\begin{equation}
\label{eq:binary_prior_objective}
\hat{x}=\arg\min_x \|Ax-b\|^2 - \log P(x),
\end{equation}
where $P(x)$ is a likelihood function measuring how likely $x$ is under the target distribution.



One way to incorporate the distributional prior in Eq.~\eqref{eq:binary_prior_objective} is to estimate $P(x)$ from $X$ and solve the resulting regularized optimization problem. A more effective approach is to perform conditional generation. We design a plug-and-play framework that combines a pretrained facial-image diffusion model with the target biometric feature extractor, allowing the generation process to be conditioned directly on constraints in the biometric feature space.

Specifically, we use a pretrained Denoising Diffusion Probabilistic Model (DDPM) $G$ that generates a facial image from an initial noise sample:
\begin{equation}
    I(z) = G(z),
    \qquad
    z \sim \mathcal{N}(0,I).
\end{equation}
The pretrained DDPM operates entirely in the image space and is not trained to generate biometric templates. Our approach applies the differentiable target feature extractor $F$ to each generated facial image, producing its corresponding biometric template:
\begin{equation}
    x(z)
    =
    F\bigl(I(z)\bigr)
    =
    F\bigl(G(z)\bigr).
\end{equation}

During sampling, we guide the image-generation process through the biometric feature space by minimizing the conditioning loss
\begin{equation}
    \mathcal{L}_{\mathrm{cond}}(z)
    =
    \left\|
        A \cdot F\bigl(G(z)\bigr) - b
    \right\|_2^2,
\end{equation}
where $A$ and $b$ encode the constraints derived from the observed authentication feedback. Equivalently, the guided sampling process seeks
\begin{equation}
    z^{*}
    =
    \arg\min_{z}
    \left\|
        A \cdot F\bigl(G(z)\bigr) - b
    \right\|_2^2.
\end{equation}
The reconstructed biometric template is then
\begin{equation}
    \hat{x}
    =
    F\bigl(G(z^{*})\bigr),
\end{equation}
and approximately satisfies
\begin{equation}
    A\hat{x} \approx b.
\end{equation}
Thus, the pretrained DDPM provides a facial-image prior, while the differentiable feature extractor enables us to guide image generation according to constraints in the target biometric feature space. This restricts the recovered template to the feature space distribution and leads to a more accurate reconstruction. The prior-based solver is summarized in Algorithm~\ref{alg:posterior_sampling}. A detailed implementation-level version of the algorithm is provided in Appendix~\ref{app:diffusion_solver_details}.

\begin{algorithm}[t] \caption{Constraint-guided diffusion reconstruction} \label{alg:posterior_sampling} \begin{algorithmic}[1] \REQUIRE DDPM noise predictor $\epsilon_\theta(\cdot,i)$; diffusion steps $N$; coefficients $\{\alpha_i,\bar{\alpha}_i\}_{i=1}^{N}$; feature extractor $F(\cdot)$; constraints $(A,b)$; guidance scales $\{\zeta_i\}_{i=1}^{N}$. \STATE Initialize $x_N\sim\mathcal{N}(0,I)$ \FOR{$i=N$ \textbf{to} $1$} \STATE $\displaystyle \hat{x}_0\gets \frac{x_i-\sqrt{1-\bar{\alpha}_i}\, \epsilon_\theta(x_i,i)} {\sqrt{\bar{\alpha}_i}}$ \STATE $\displaystyle \mathcal{L}_{\mathrm{con}} \gets \left\|A\cdot F(\hat{x}_0)-b\right\|_2^2$ \STATE $\displaystyle x_{i-1}\gets \frac{1}{\sqrt{\alpha_i}} \left( x_i- \frac{1-\alpha_i}{\sqrt{1-\bar{\alpha}_i}} \epsilon_\theta(x_i,i) \right) - \zeta_i\nabla_{x_i}\mathcal{L}_{\mathrm{con}}$ \ENDFOR \STATE \textbf{return} $F(\hat{x}_0)$ \end{algorithmic} \end{algorithm}

\section{Experimental Evaluation}
\label{sec:results}

Our experimental evaluation is guided by the following research questions:
\begin{itemize}[leftmargin=15pt]
    \item \textbf{RQ1:} Can protected biometric templates be reconstructed from binary authentication feedback alone?

    \item \textbf{RQ2:} Does the proposed query-efficient score-based attack achieve lower reconstruction error than prior score-based attacks for every query budget?

    \item \textbf{RQ3:} Does the proposed prior-based solver effectively approximate the distribution-constrained objective in Eq.~\eqref{eq:dist_equation} and achieve improved reconstruction when the attacker cannot choose the breaking-set (e.g., when the breaking-set is sampled from a random distribution)?

    \item \textbf{RQ4:} How much does the proposed query-efficient binary-feedback attack reduce the number of authentication queries?

\end{itemize}

\definecolor{bestrow}{RGB}{245,230,230}
\begin{figure}[htbp]
    \centering
    \includegraphics[width=\linewidth]{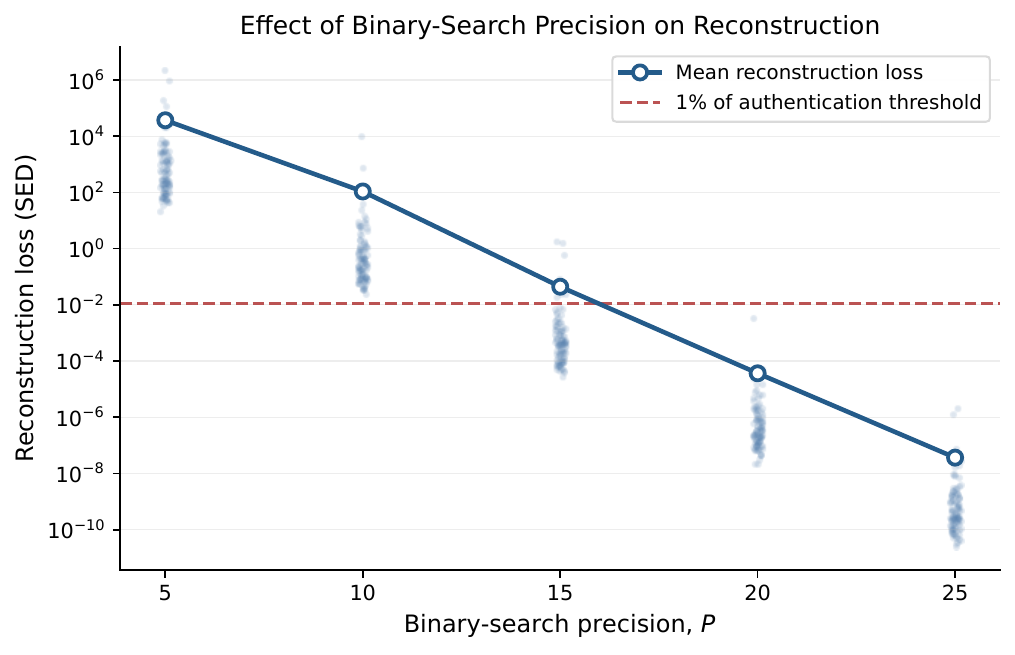}
    \caption{Effect of the binary-search precision parameter $P$ on reconstruction loss. Each point represents a target template, and the solid line shows the mean reconstruction loss. The dashed line indicates 1\% of the authentication threshold.}
    \label{fig:p_ablation}
\end{figure}

\begin{table*}[t]
    \centering
    \caption{Comparison of the evaluated reconstruction methods}
    \label{tab:method_query_comparison}
    \renewcommand{\arraystretch}{1.15}
    \resizebox{\textwidth}{!}{
    \begin{tabular}{l l c}
        \toprule
        \textbf{Experimental Setting} &
        \textbf{Method} &
        \textbf{Queries to Reconstruction Error $< 1\%$} \\
        \midrule

        \multirow{3}{*}{\shortstack[l]{Score-releasing\\with input selection}}
        & Hill climbing~\cite{6994220}
        & $>500$ \\

        & Algebraic reconstruction with input selection~\cite{4359297}
        & $\approx 60$ \\

        & \cellcolor{bestrow}\textbf{PCA-based input selection (Ours)}
        & \cellcolor{bestrow}$\approx$~\textbf{50} \\

        \midrule

        \multirow{4}{*}{\shortstack[l]{Score-releasing\\without input selection}}
        & Lagrange-based reconstruction~\cite{10449211}
        & $>500$ \\

        & Algebraic reconstruction without input selection~\cite{4359297}
        & $>500$ \\

        & GMM-prior reconstruction (Ours -- Baseline)
        & $>500$ \\

        & \cellcolor{bestrow}\textbf{Diffusion-prior reconstruction (Ours)}
        & \cellcolor{bestrow}$\approx$~\textbf{90} \\

        \midrule

        \multirow{3}{*}{Binary feedback}
        & Averaging accepted false matches (Ours -- Baseline)
        & Not achieved \\

        & Geometric attack with algebraic solver (Ours)
        & $>10,000$ \\

        & \cellcolor{bestrow}\textbf{Geometric attack with diffusion solver (Ours)}
        & \cellcolor{bestrow}$\approx$~\textbf{2,000} \\

        \bottomrule
    \end{tabular}
    }
\end{table*}

\subsection{Experimental Setup}
\label{sec:setup}
For feature extraction, we use standard algorithms and implementations, specifically ArcFace\footnote{ArcFace implementation: \url{https://github.com/peteryuX/arcface-tf2}.}~\cite{Deng_2019_CVPR} and FaceNet\footnote{FaceNet implementation: \url{https://github.com/timesler/facenet-pytorch}.}~\cite{Schroff_2015_CVPR}, which are also used in prior protected-recognition systems~\cite{8698601,8897272,9794523}. We determine the system thresholds for False Match Rates (FMRs) of 1.0\% and 0.1\%.

We use the standard LFW~\cite{huang2008labeled} and celebA~\cite{Liu_2015_ICCV} benchmarks for face verification. Both contains more than 10,000 facial images and 1,000 identities. For each benchmark, we randomly selected 300 identities  as the target users of the system.

We evaluate reconstruction accuracy using the template reconstruction loss, defined as the distance between the reconstructed and target templates under the corresponding matching metric (SED or cosine similarity). We assess query efficiency by measuring reconstruction loss as a function of the number of authentication queries, including ablations examining the effect of the binary-search precision parameter $P$. Additionally, we report the percentage of targets achieving reconstruction loss below 1\% of the corresponding authentication threshold and the number of queries required to reach this near-perfect reconstruction criterion.

Table~\ref{tab:method_query_comparison} summarizes all the evaluated methods.

\subsection{Binary Reconstruction Evaluation}
\begin{tcolorbox}[
  colback=blue!5,
  colframe=blue!35,
  arc=6pt,
  boxrule=0.6pt,
  left=10pt,
  right=10pt,
  top=8pt,
  bottom=8pt,
  before skip=8pt,
  after skip=8pt,
  title=\textbf{Key Takeaway:},
  fonttitle=\bfseries,
  coltitle=black
]
Binary authentication feedback alone enables highly accurate template reconstruction, with reconstruction loss below 1\% in 97.6\%--100\% of cases, answering RQ1.
\end{tcolorbox}
\textit{Target Systems}: 
We evaluate binary-only template reconstruction attacks under two types of biometric decision rules: systems based on squared Euclidean distance (SED) and systems based on cosine similarity. 
We use the feature extractors and thresholds described in Section~\ref{sec:setup}. 
Facial templates of enrolled users are reconstructed using our method, as detailed in Section~\ref{sec:binary_reconstruction}, where $d=512$ is the feature-vector dimension for both ArcFace and FaceNet. 
For comparison, we use the Baseline approach described in Section~\ref{sec:binary_baseline}.

\textit{Implementation Details}: 
we set $P=20$ and for a fair comparison, the Baseline approach is allotted the same number of authentication attempts as our method in each setting. 
In the Baseline approach, the breaking-set consists of all available images that do not belong to the target identity. 
The total number of authentication attempts is therefore identical for both methods within each setting. 

\textit{Results}: 
Reconstruction accuracy is evaluated by measuring the percentage of target templates whose reconstruction loss is below 1\% of the corresponding system decision threshold. As shown in Table~\ref{tab:loss_below_1pct_all_}, our method achieves highly accurate reconstruction across both datasets, feature extractors, decision rules, and FMR operating points. Across all evaluated settings, between 97.6\% and 100\% of the reconstructed templates produced by our method fall below this stringent 1\% threshold. In contrast, the Baseline approach almost never reaches this level of accuracy, achieving 0\% in most settings and at most 4.0\% in the best case. These results demonstrate that binary authentication feedback can be sufficient for highly accurate template reconstruction when the attack exploits the geometry of the decision boundary.

To examine the effect of the binary-search precision parameter $P$, we additionally evaluate the reconstruction loss for different values of $P$. 
As shown in Figure~\ref{fig:p_ablation}, reconstruction improves rapidly as $P$ increases, approaching perfect reconstruction up to numerical precision. At $P=20$, almost all targets achieve reconstruction loss below $1\%$ of the authentication threshold.
\begin{table*}[htbp]
    \centering
    \caption{Percentage of targets with reconstruction loss below 1\% of the corresponding system threshold. For cosine systems, the threshold comparison uses normalized angular distance.}
    \label{tab:loss_below_1pct_all_}
    \resizebox{0.88\textwidth}{!}{%
    \begin{tabular}{
        lll
        c >{\columncolor{bestrow}}c
        c >{\columncolor{bestrow}}c
    }
        \toprule
        \textbf{Decision Rule}
        & \textbf{Feature Extractor}
        & \textbf{FMR}
        & \multicolumn{2}{c}{\textbf{LFW}}
        & \multicolumn{2}{c}{\textbf{CelebA}} \\
        \cmidrule(lr){4-5}
        \cmidrule(lr){6-7}
        &
        &
        &
        \textbf{Baseline}
        & \textbf{Ours}
        & \textbf{Baseline}
        & \textbf{Ours} \\
        \midrule
        \multirow{4}{*}{SED} & \multirow{2}{*}{ArcFace} & 1\% & 0.0\% & \textbf{99.3\%} & 0.0\% & \textbf{99.3\%} \\
         &  & 0.1\% & 3.5\% & \textbf{99.6\%} & 0.0\% & \textbf{100.0\%} \\
        \cmidrule(lr){2-7}
         & \multirow{2}{*}{FaceNet} & 1\% & 0.0\% & \textbf{99.3\%} & 0.0\% & \textbf{99.7\%} \\
         &  & 0.1\% & 0.0\% & \textbf{99.3\%} & 0.0\% & \textbf{100.0\%} \\
        \midrule
        \multirow{4}{*}{Cosine} & \multirow{2}{*}{ArcFace} & 1\% & 0.0\% & \textbf{98.6\%} & 0.0\% & \textbf{100.0\%} \\
         &  & 0.1\% & 4.0\% & \textbf{98.8\%} & 0.0\% & \textbf{100.0\%} \\
        \cmidrule(lr){2-7}
         & \multirow{2}{*}{FaceNet} & 1\% & 0.0\% & \textbf{99.5\%} & 0.0\% & \textbf{97.6\%} \\
         &  & 0.1\% & 0.0\% & \textbf{98.2\%} & 0.0\% & \textbf{98.9\%} \\
        \bottomrule
    \end{tabular}%
    }
\end{table*}

\begin{figure}[htbp]
    \centering
    \includegraphics[width=0.48\linewidth]{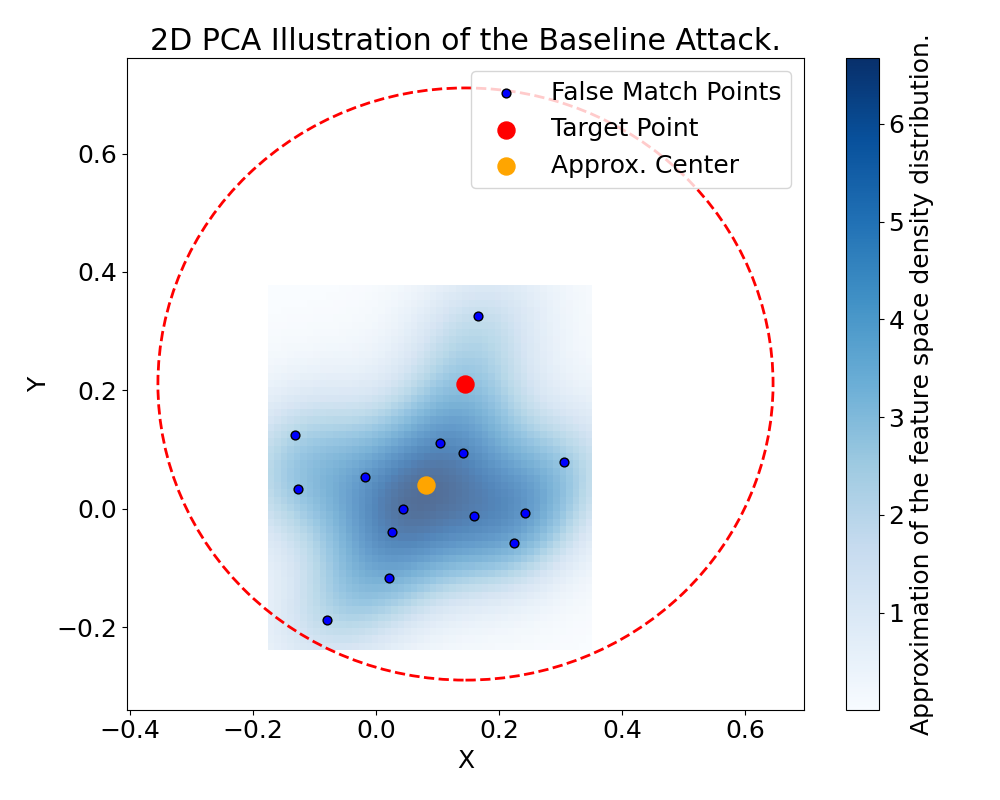}
     \includegraphics[width=0.48\linewidth]{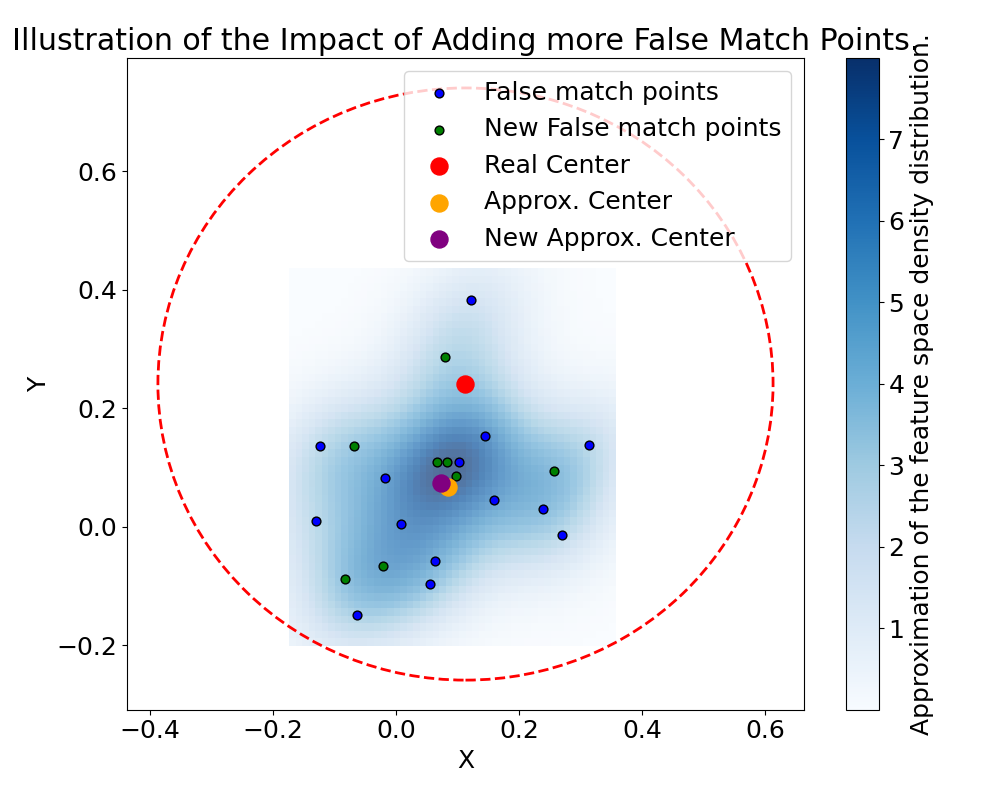}
    \caption{Illustration of the baseline attack using 2D PCA: left -- the approximation of the feature space density distribution within the match space of the target template; right -- Illustration of the impact of adding more false match points.}
    \label{fig:pca1}
\end{figure}
furthermore, the results also highlight a major limitation of the Baseline approach, even when many false-match points are available, the reconstruction quality remains limited and cannot achieve perfect reconstruction. While it may seem intuitive that averaging an increasing number of false match points would converge to the real template, this only holds if those points are uniformly distributed around the center of the sphere. However, this assumption does not hold in practice. The false match points are not only constrained to the sphere but are also distributed according to the underlying structure of the feature space. Figure \ref{fig:pca1}-left illustrates an approximation of the feature space density distribution within the match space of the target template in 2D using PCA. In Figure \ref{fig:pca1}-right, it can be seen that new false match points fall within the same feature space distribution, even though these new false match points are from different images of different identities. The improvement of the baseline approach over time can be seen in Figure~\ref{fig:yes_no_results}.

\begin{figure*}[htbp]
    \centering
    \includegraphics[width=0.7\textwidth]{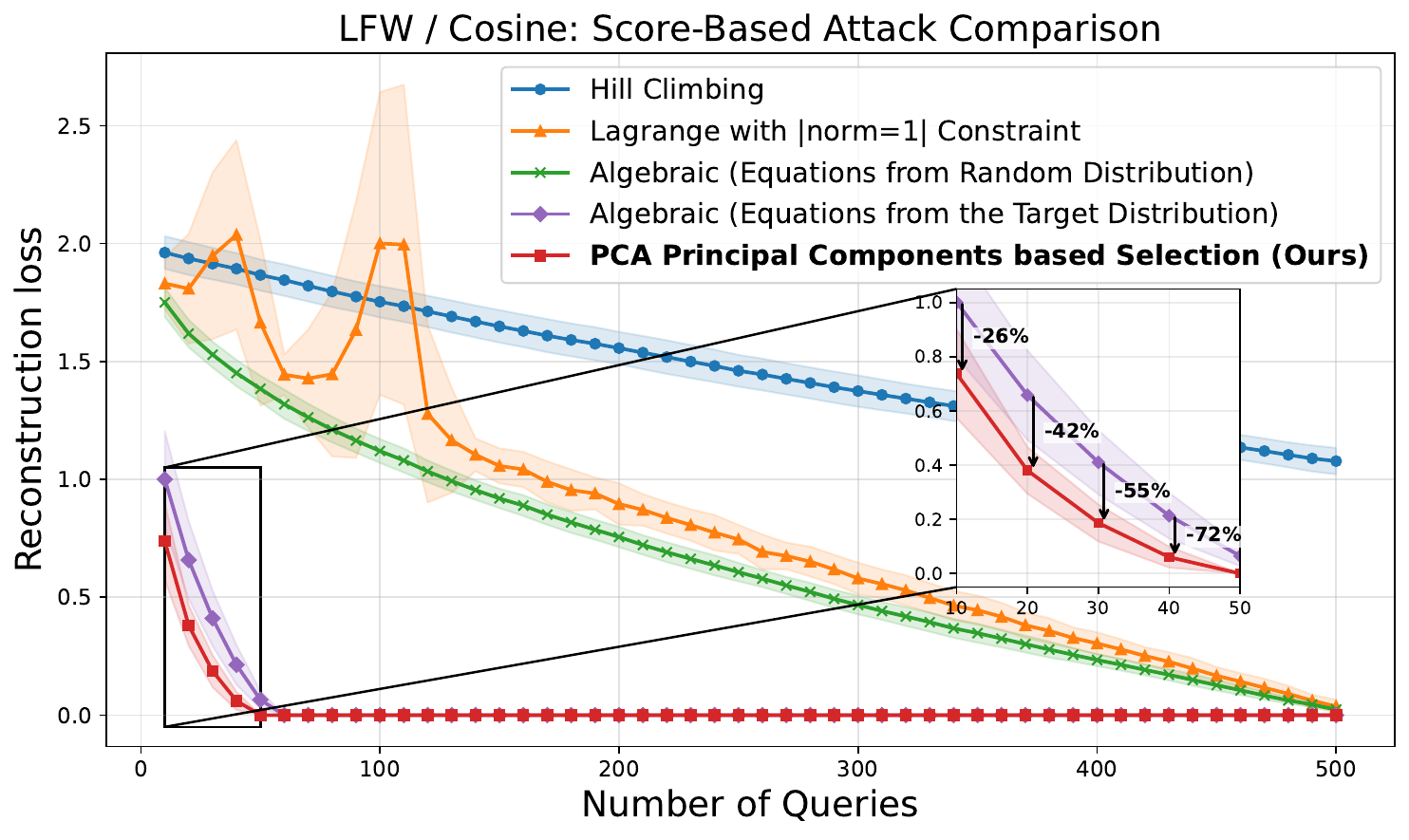}
    \captionsetup{justification=centering}
    \caption{Reconstruction loss as a function of the number of queries for different reconstruction strategies that use input selection.}
    \label{fig:efficient_scores}
\end{figure*}

\begin{figure*}[ht]
    \centering
    \includegraphics[width=0.7\textwidth]{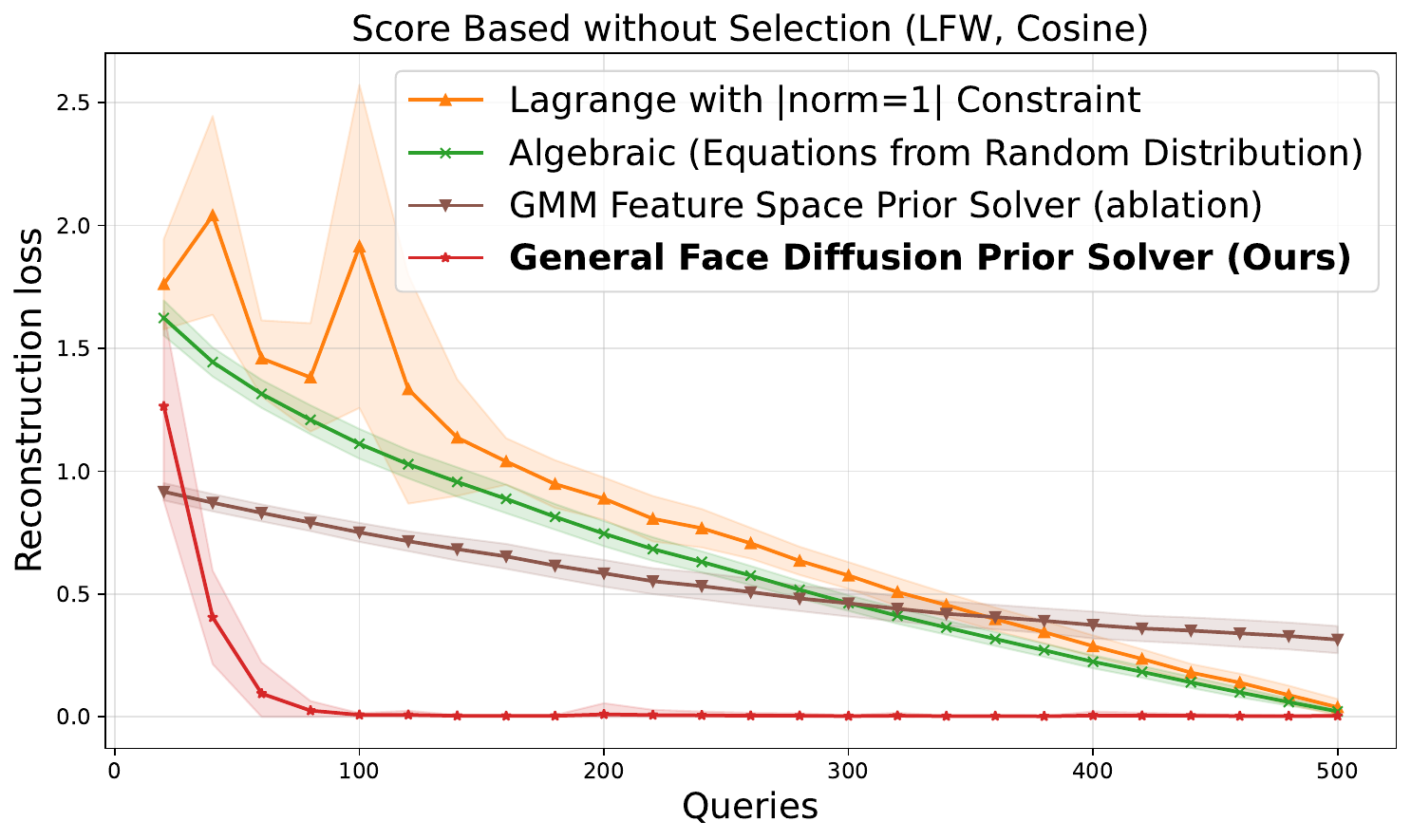}
    \captionsetup{justification=centering}
    \caption{Reconstruction loss as a function of the number of queries/equations for different reconstruction strategies that do not use input selection.}
    \label{fig:no_selection_scores}
\end{figure*}

\subsection{Evaluation of Query-Efficient Reconstruction from Similarity Scores}
\label{sec:eval_efficent_score}
\begin{tcolorbox}[
  colback=blue!5,
  colframe=blue!35,
  arc=6pt,
  boxrule=0.6pt,
  left=10pt,
  right=10pt,
  top=8pt,
  bottom=8pt,
  before skip=8pt,
  after skip=8pt,
  title=\textbf{Key Takeaway:},
  fonttitle=\bfseries,
  coltitle=black
]
Our PCA-based input-selection method consistently achieves the lowest reconstruction error across all evaluated query budgets, addressing RQ2.
\end{tcolorbox}

\textit{Compared Methods}. We evaluate the query-efficient breaking-set selection strategy introduced in Section~\ref{sec:query_efficient_scores}. We compare it against several score-based reconstruction baselines: (i) hill climbing, following prior score-leakage attacks~\cite{6994220}; (ii) a Lagrange-based reconstruction method with a unit-norm constraint, $\|x\|=1$ for biometric template reconstruction presented in\footnote{https://github.com/aminabassit/tra-he-btps.}~\cite{10449211}; (iii) an algebraic reconstruction method presented in~\cite{4359297}, whose equations are obtained from samples drawn from a random distribution; and (iv) and the same method presented in~\cite{4359297} where the equations are obtained from samples drawn from the target feature distribution. All methods are evaluated under multiple query budgets, ranging from a small number of queries to $500$ queries. 
Additional implementation details, including hyperparameter settings for our method and the compared baselines, are provided in Appendix~\ref{app:extended_implementation_details}.

\textit{Results}. Figure~\ref{fig:efficient_scores} shows the average reconstruction loss as a function of the number of available queries. Our PCA-based input-selection method achieves the lowest reconstruction loss for every evaluated query budget, directly answering RQ2. The loss decreases rapidly with only a few dozen queries, and the method reaches perfect reconstruction, up to numerical precision, after roughly $50$ queries. The results also show that query selection is the key factor behind query-efficient score-based reconstruction. In particular, the same algebraic solver behaves very differently depending on how the equations are selected. When the equations are constructed from templates sampled from the target feature-space distribution, the reconstruction loss decreases rapidly and approaches numerical precision. In contrast, when the equations are constructed from random vectors outside this distribution, the loss remains much higher, around $0.14$ even at the largest evaluated budget of $500$ queries. This gap shows that distribution-aware queries provide substantially more informative geometric constraints on the unknown template than random equations. Compared with simple random selection from the target feature-space distribution, our PCA-based input-selection method achieves a lower reconstruction loss at every evaluated query budget, corresponding to an improvement of between $26\%$ and $72\%$. Compared with prior baselines that do not exploit the target feature-space distribution, Our method requires approximately $5\times$ fewer queries to reach numerical-precision reconstruction within the evaluated range. Additional results for all benchmark and decision-rule combinations are provided in Appendix\ref{app:scores_with}, further supporting these findings.

\subsection{Evaluation of Query-Efficient Reconstruction from Similarity Scores Without Input Selection}
\begin{tcolorbox}[
  colback=blue!5,
  colframe=blue!35,
  arc=6pt,
  boxrule=0.6pt,
  left=10pt,
  right=10pt,
  top=8pt,
  bottom=8pt,
  before skip=8pt,
  after skip=8pt,
  title=\textbf{Key Takeaway:},
  fonttitle=\bfseries,
  coltitle=black
]
When input selection is unavailable (e.g., when the breaking set is sampled from a random distribution), our proposed prior-based solver improves reconstruction over prior solvers and achieves performance comparable to the case where the breaking set is sampled directly from the target distribution, indicating that it effectively approximates Eq.~\eqref{eq:dist_equation}, addressing RQ3.
\end{tcolorbox}

\noindent\textbf{Setup:}
As discussed in Section~\ref{sec:query_efficient_binary}, when input selection is unavailable, reconstruction performance primarily depends on the solver's ability to extract information from a fixed set of equations. To evaluate this setting, we construct the breaking-set by sampling from a random distribution and compare our method against the baselines evaluated in Section~\ref{sec:eval_efficent_score}.

The results in Figure~\ref{fig:no_selection_scores} show that the diffusion-based approach consistently outperforms both classical algebraic methods and the Gaussian prior approach. Notably, the diffusion-based solver achieves performance close to the ideal setting where input selection is available, despite operating without access to such control. This indicates that the proposed prior-based solver effectively approximates the distribution-constrained objective in Eq.~\eqref{eq:dist_equation}, as its reconstruction quality is close to the algebraic solution obtained using equations sampled from the target distribution. Additional results for every benchmark and decision-rule combination are provided in Appendix\ref{app:scores_without}, further supporting the findings of this work.

\subsection{Query-Efficient Binary Reconstruction Evaluation}
\label{sec:eval_efficent_score}
\begin{tcolorbox}[
  colback=blue!5,
  colframe=blue!35,
  arc=6pt,
  boxrule=0.6pt,
  left=10pt,
  right=10pt,
  top=8pt,
  bottom=8pt,
  before skip=8pt,
  after skip=8pt,
  title=\textbf{Key Takeaway:},
  fonttitle=\bfseries,
  coltitle=black
]
The proposed query-efficient binary-feedback attack substantially reduces the number of authentication queries. In particular, by integrating the diffusion-based solver, the attack reaches perfect reconstruction using approximately $5\times$ fewer queries than the algebraic solver, directly addressing RQ4.
\end{tcolorbox}

We evaluate whether the diffusion-based prior can reduce the query complexity of the binary-feedback reconstruction attack. Both the algebraic and diffusion-based variants use the same binary-search procedure to obtain points on the authentication boundary; they differ only in how the resulting constraints are used to reconstruct the target template. The algebraic solver requires sufficiently many independent boundary constraints to determine the template directly, whereas the diffusion-based solver exploits the facial-image prior to recover an accurate template from a substantially smaller number of constraints. Since constructing each additional boundary constraint requires $P$ authentication queries, reducing the number of required constraints directly reduces the total query cost.

\begin{figure*}[ht]
    \centering
    \includegraphics[width=0.7\textwidth]{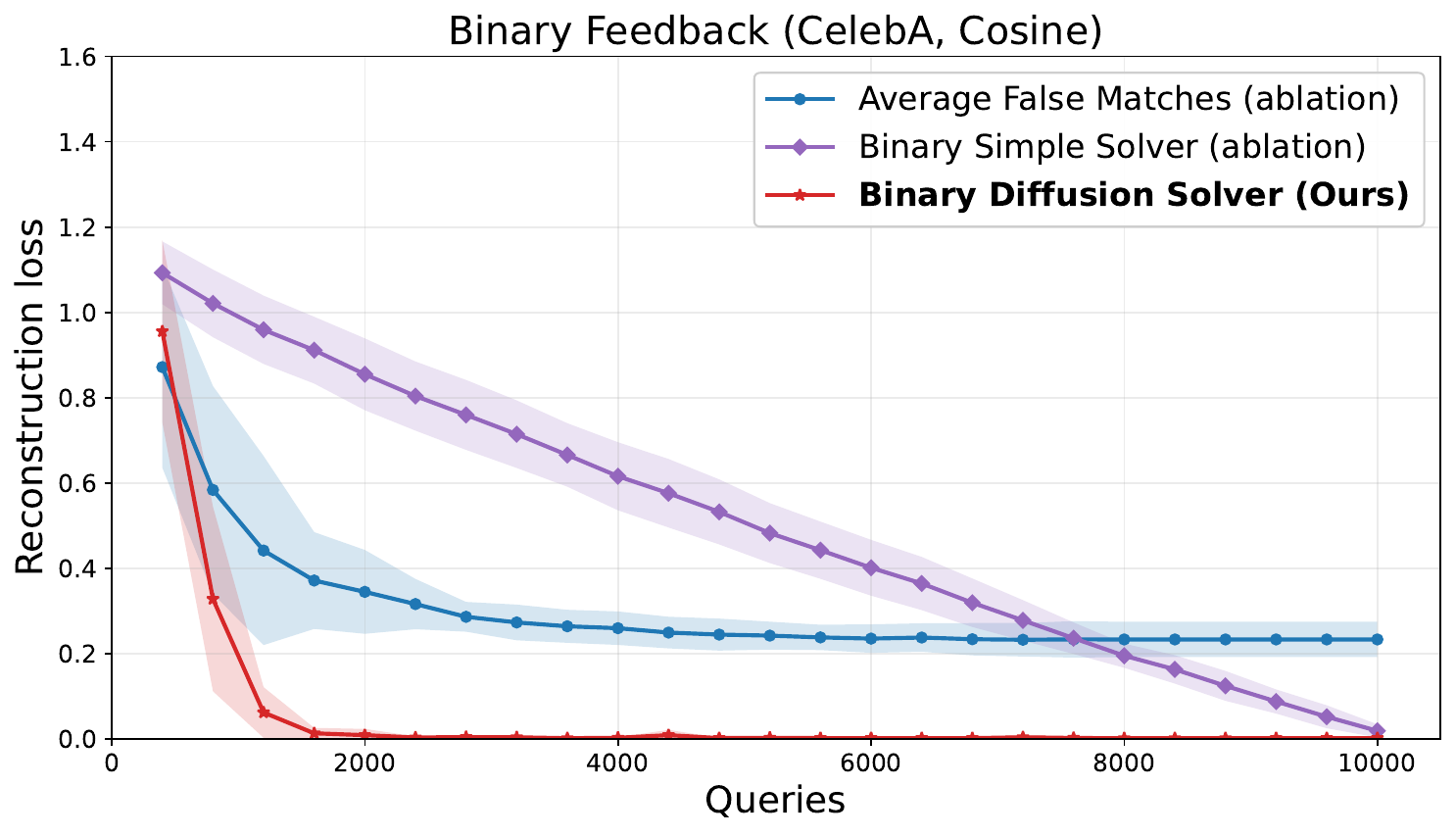}
    \captionsetup{justification=centering}
    \caption{Reconstruction loss as a function of the number of queries for different binary-feedback reconstruction strategies.}
    \label{fig:yes_no_results}
\end{figure*}

Figure~\ref{fig:yes_no_results} reports the reconstruction loss as a function of the total number of authentication queries. Across the evaluated query budgets, the diffusion-based solver achieves the lowest reconstruction loss and converges considerably faster than the algebraic solver. In particular, the algebraic solver requires approximately $10{,}000$ queries to achieve perfect reconstruction up to numerical precision, whereas the diffusion-based solver reaches the same reconstruction accuracy using approximately $5\times$ fewer authentication queries, directly answering RQ4.

In contrast, the baseline averaging attack improves only during the initial queries and then quickly saturates at a non-negligible reconstruction loss. Even when additional false-match templates are collected, averaging them does not recover the target template accurately because these points are not distributed symmetrically around the target. Overall, these results show that combining binary-search-based reconstruction with the diffusion-based prior solver provides a significant improvement in query efficiency, enabling perfect reconstruction with far fewer authentication attempts.
Additional results for every benchmark and decision-rule combination are provided in Appendix\ref{app:binary}, further supporting the findings of this work.
\section{Security Implications and Discussion}

\paragraph{Limitations of Cryptographic Protection}
Cryptographic techniques such as FHE are designed to protect templates at rest and during computation. However, our findings demonstrate that these protections do not prevent information leakage through repeated interactions with the system. In particular, the authentication interface itself becomes a side channel, enabling an adversary to gradually recover the protected template through carefully designed queries.

\paragraph{Role of Feature Space Geometry}
Our analysis highlights the importance of the geometric structure of deep feature spaces. In SED systems, binary authentication defines a sphere around the target template. In cosine similarity systems, the normalization constraint induces a spherical geometry and the acceptance region becomes a cone on the unit hypersphere. In both cases, the geometry of the decision boundary can be exploited for reconstruction. Furthermore, incorporating distribution-aware priors dramatically improves reconstruction, indicating that the structure of the feature space can be exploited by adversaries.

\paragraph{Implications for System Design}
Our findings show that encrypting biometric templates and restricting the system output to a similarity score or a binary decision are not sufficient when an attacker can issue many adaptive queries. Secure biometric system design must therefore address the cumulative information leaked through repeated authentication attempts. Possible defenses include enforcing strict query limits, detecting query sequences that search for the decision boundary, randomizing or perturbing the returned scores and decisions, and introducing controlled uncertainty into the matching threshold. In addition, future biometric representations should be designed to preserve recognition accuracy while reducing the amount of geometric information that can be exploited for template reconstruction.

\paragraph{Limitations of This Work}
Our attacks assume black-box access to the authentication system and, in some cases, access to a dataset generated by the same feature extractor. While these assumptions are realistic in many practical scenarios, further work is needed to evaluate robustness under stricter constraints, mismatched distributions, and systems with active query-detection mechanisms.

Overall, our results demonstrate that current privacy-preserving biometric systems may provide weaker protection than previously believed, particularly against adversaries capable of exploiting query-based information leakage.

\section{Conclusion and Future Work}

In this work, we presented a unified framework for reconstructing protected biometric templates under extremely limited feedback. We demonstrated that both similarity scores and binary authentication outcomes can be exploited to recover biometric templates with high accuracy, even under query constraints.

We introduced reconstruction attacks from binary authentication feedback, covering both SED and cosine similarity systems. We further introduced query-efficient reconstruction attacks for score-releasing systems through distribution-aware input selection, and query-efficient binary reconstruction using generative priors and diffusion-based posterior sampling.

Our experimental results show that these attacks achieve low reconstruction loss while requiring substantially fewer queries than existing approaches, highlighting a critical gap between theoretical protection guarantees and practical security.

A few interesting directions for future work include extending the proposed attacks beyond face recognition to other biometric modalities, such as fingerprints, iris recognition, and voice embeddings. Another direction is to investigate whether the assumption of access to the system's exact feature extractor can be relaxed. In particular, an attacker may use a surrogate feature extractor and exploit transferability across models, potentially relying on similar feature distributions or shared representational structures, as suggested by the Platonic Representation Hypothesis~\cite{huh2024platonicrepresentationhypothesis}.

We hope this work motivates further research into the design of biometric systems that remain secure even under repeated and adaptive querying, and highlights the need for stronger guarantees beyond limiting observable outputs.

\subsection*{Acknowledgements}

The research in this paper was partially supported by the Center for Cyber, Law and Policy at the University of Haifa in conjunction with the Israel National Cyber Directorate in the Prime Minister’s Office. We would like to thank Adi Shamir and Arun Ross for the helpful discussions. We would also like to thank Marc Stevens, David Niehues, and Dennis Jackson for some interesting Twitter discussions about the mathematical problem at the heart of our algorithm.

\clearpage
\bibliographystyle{IEEEtran}
\bibliography{egbib}

\clearpage
\appendices

\section{}
\subsection{Additional Results}
\label{app:results}

\begin{figure*}[h]
    \centering
    \includegraphics[width=0.48\textwidth]{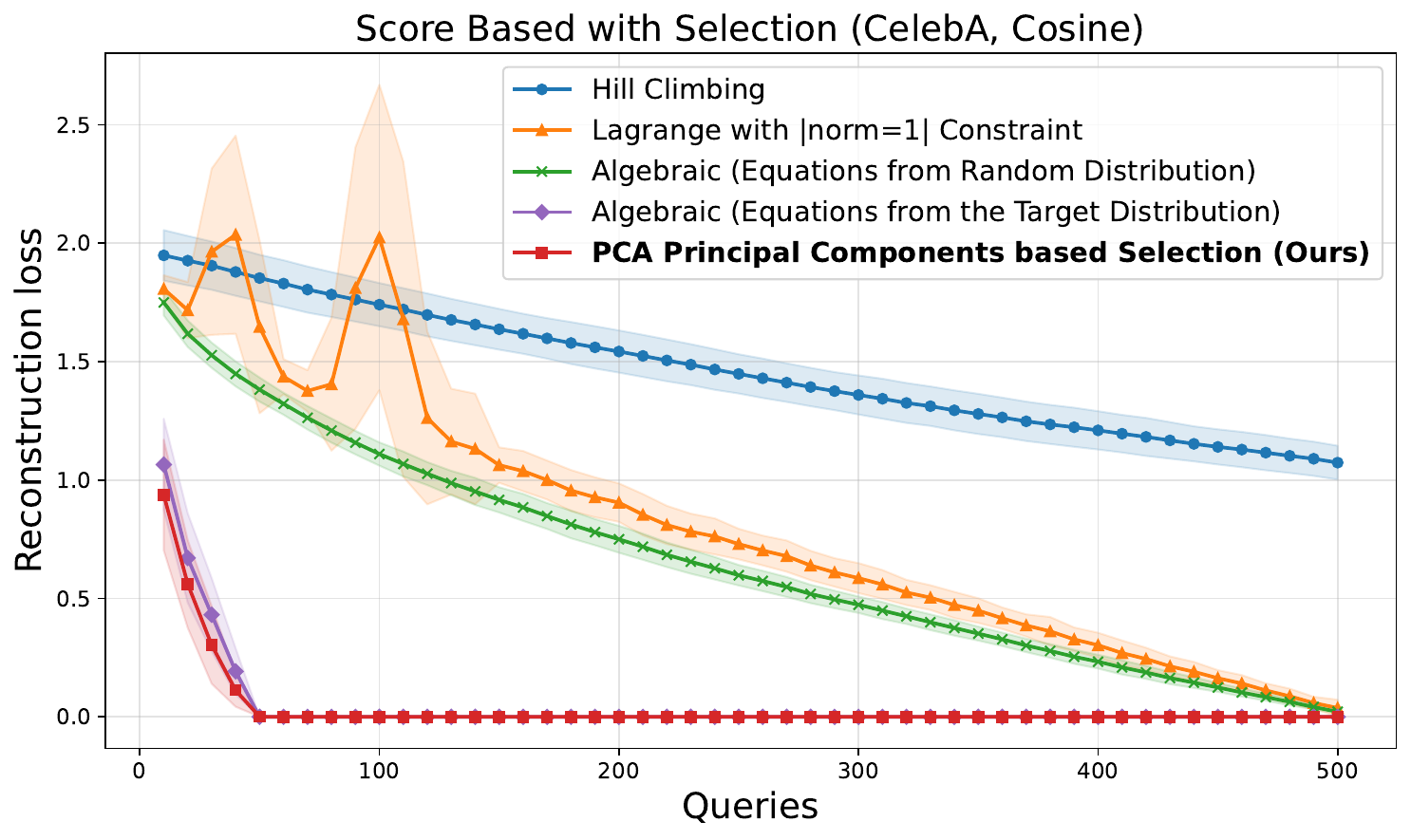}
    \hfill
    \includegraphics[width=0.48\textwidth]{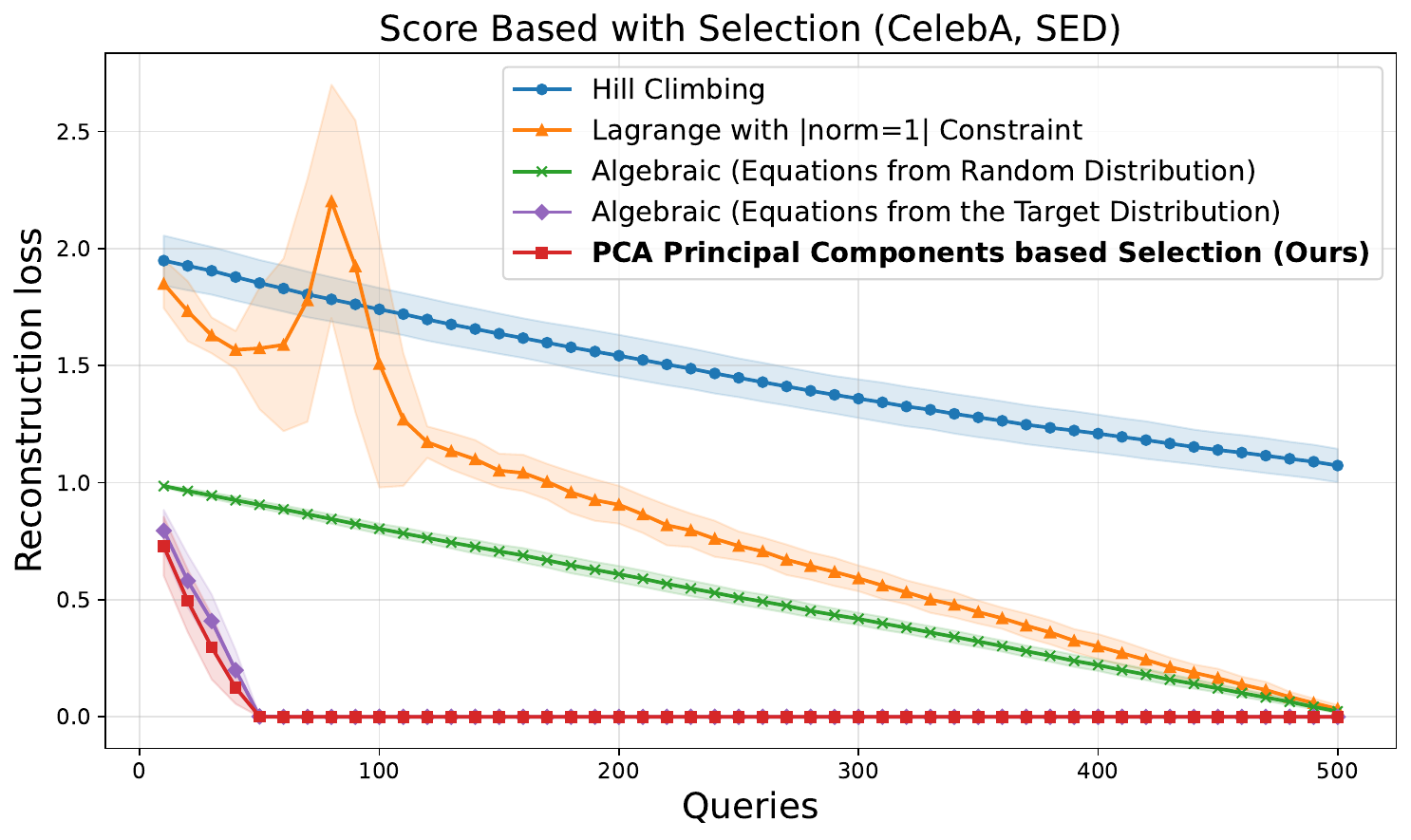}
    
    \vspace{0.5em}
    
    \includegraphics[width=0.48\textwidth]{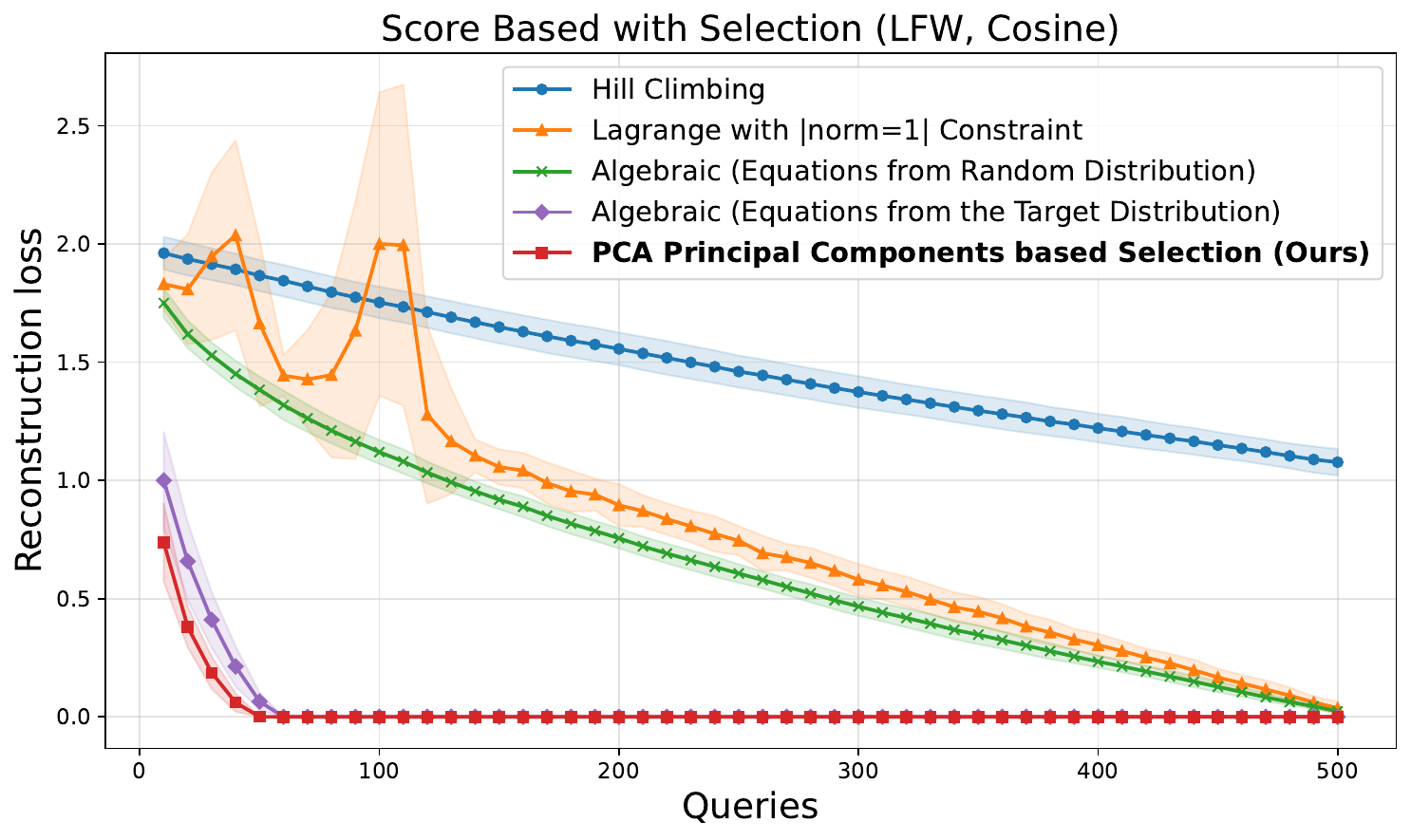}
    \hfill
    \includegraphics[width=0.48\textwidth]{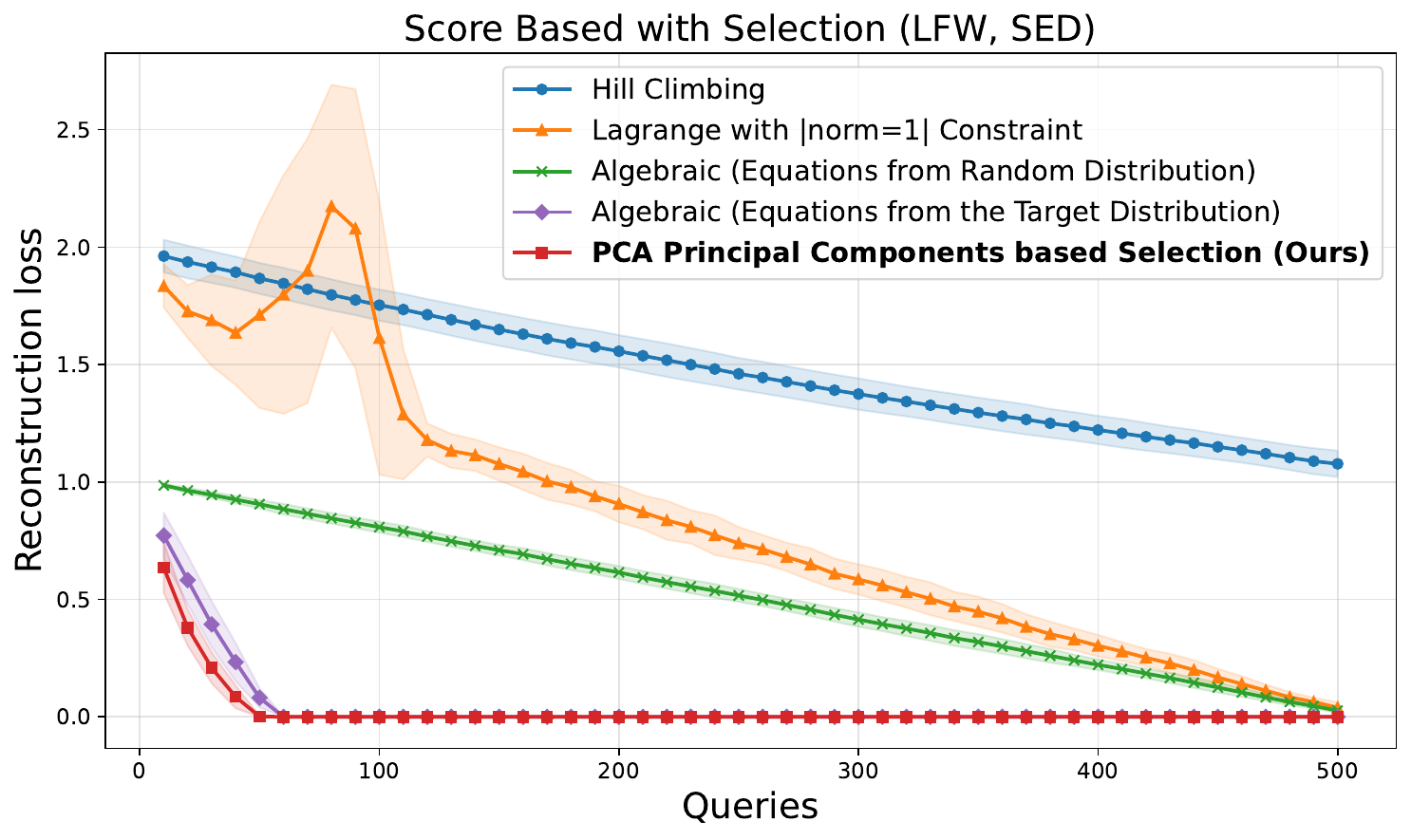}
    \caption{Additional benchmark and decision-boundary results for query-efficient reconstruction from similarity scores with input selection, across CelebA and LFW under cosine and SED similarity measures. The results are consistent with those reported in the main text and further support our findings.}
    \label{fig:appendix_score_with_selection}
\end{figure*}

\begin{figure*}[h]
    \centering
    \includegraphics[width=0.48\textwidth]{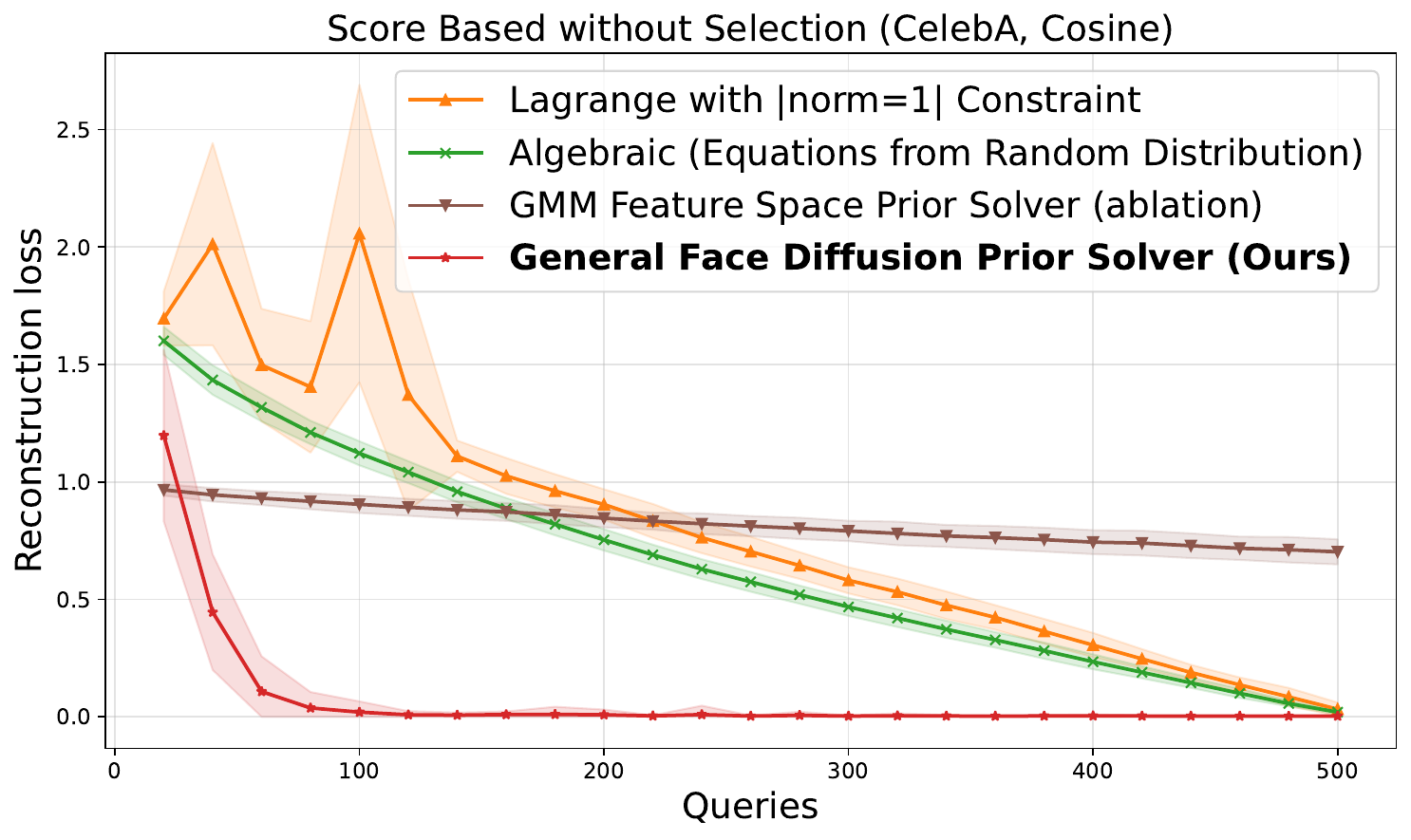}
    \hfill
    \includegraphics[width=0.48\textwidth]{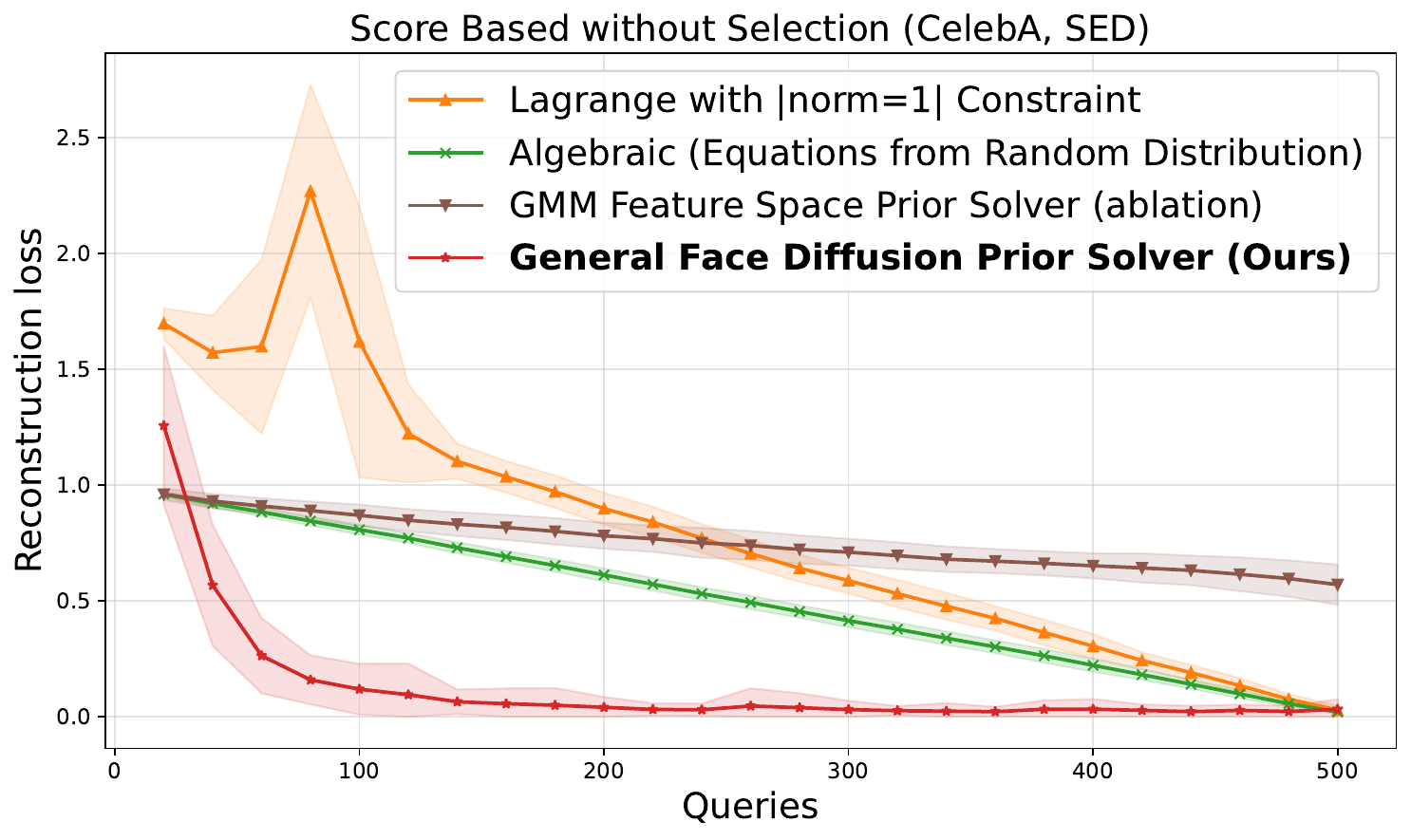}
    
    \vspace{0.5em}
    
    \includegraphics[width=0.48\textwidth]{fig/score_without_input_selection/lfw_cosine.pdf}
    \hfill
    \includegraphics[width=0.48\textwidth]{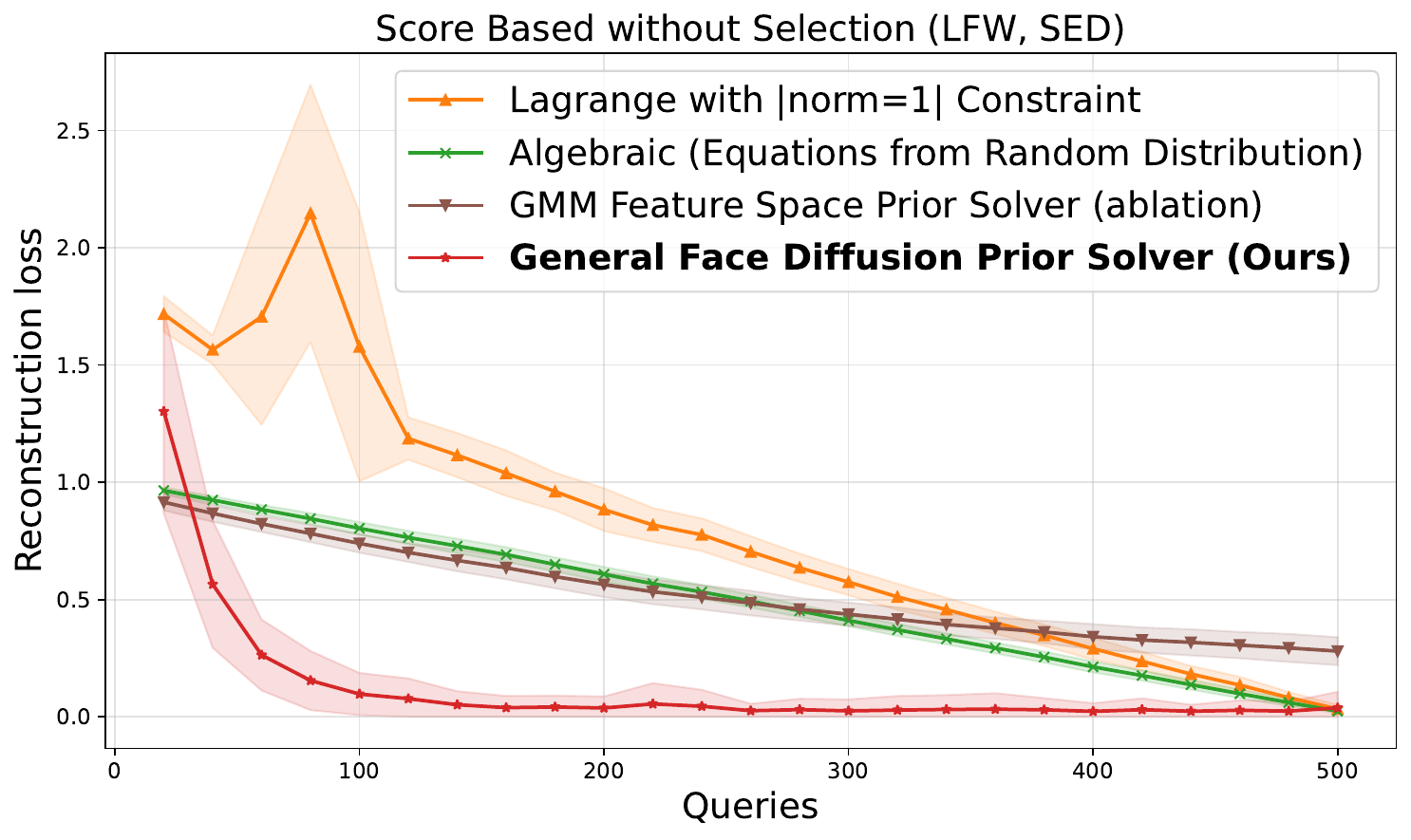}
    \caption{Additional benchmark and decision-boundary results for query-efficient reconstruction from similarity scores without input selection, across CelebA and LFW under cosine and SED similarity measures. These results are consistent with the main-text results and support the same conclusions.}
    \label{fig:appendix_score_without_selection}
\end{figure*}

We additionally report results for every combination of benchmark and decision rule to further validate the consistency of our findings across experimental settings. Across all settings, these additional results are consistent with the quantitative results presented in the main text and further reinforce our conclusions.

\begin{figure*}[h]
    \centering
    \includegraphics[width=0.48\textwidth]{fig/binary_score/celebA_cosine.pdf}
    \hfill
    \includegraphics[width=0.48\textwidth]{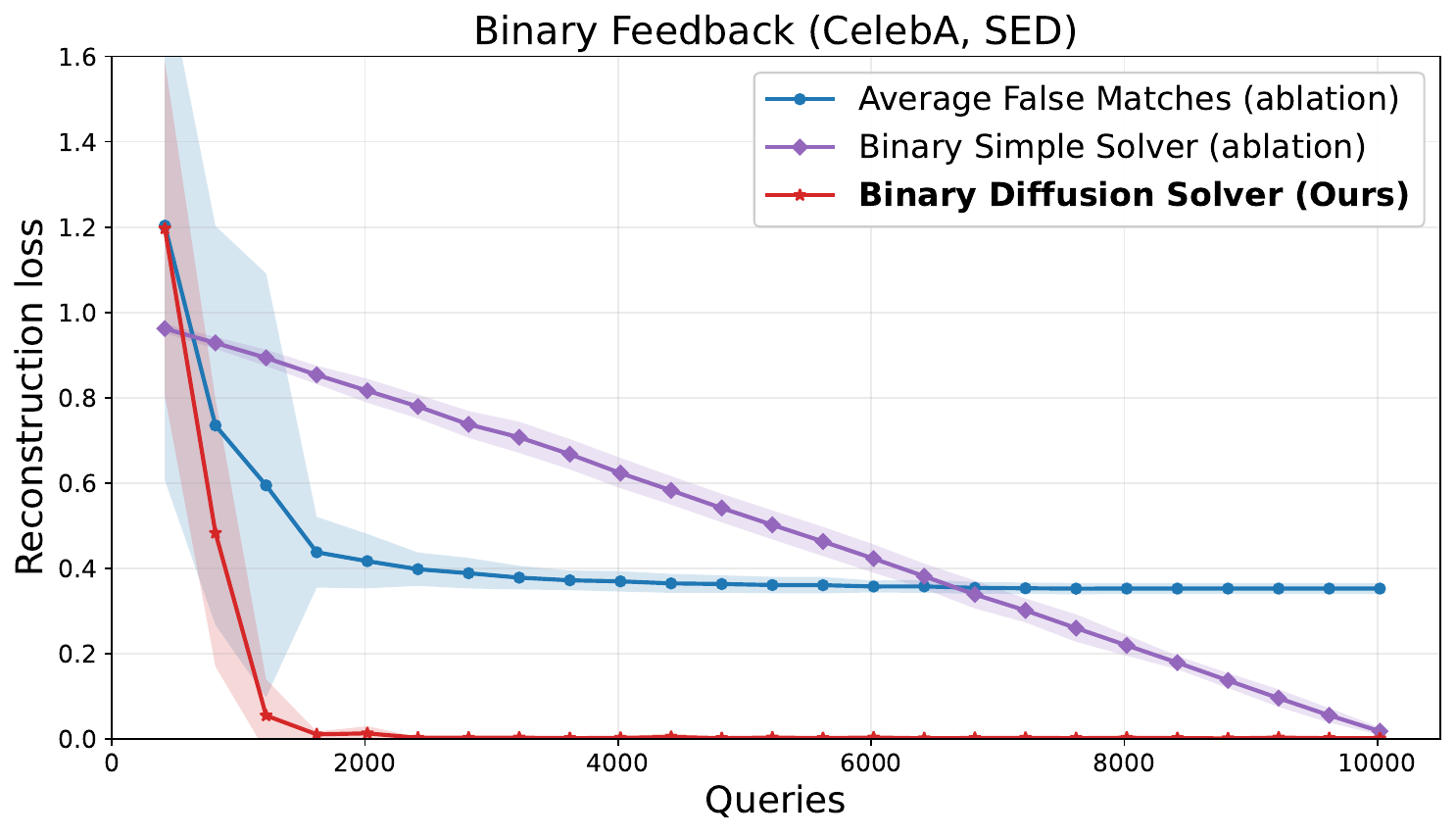}
    
    \vspace{0.5em}
    
    \includegraphics[width=0.48\textwidth]{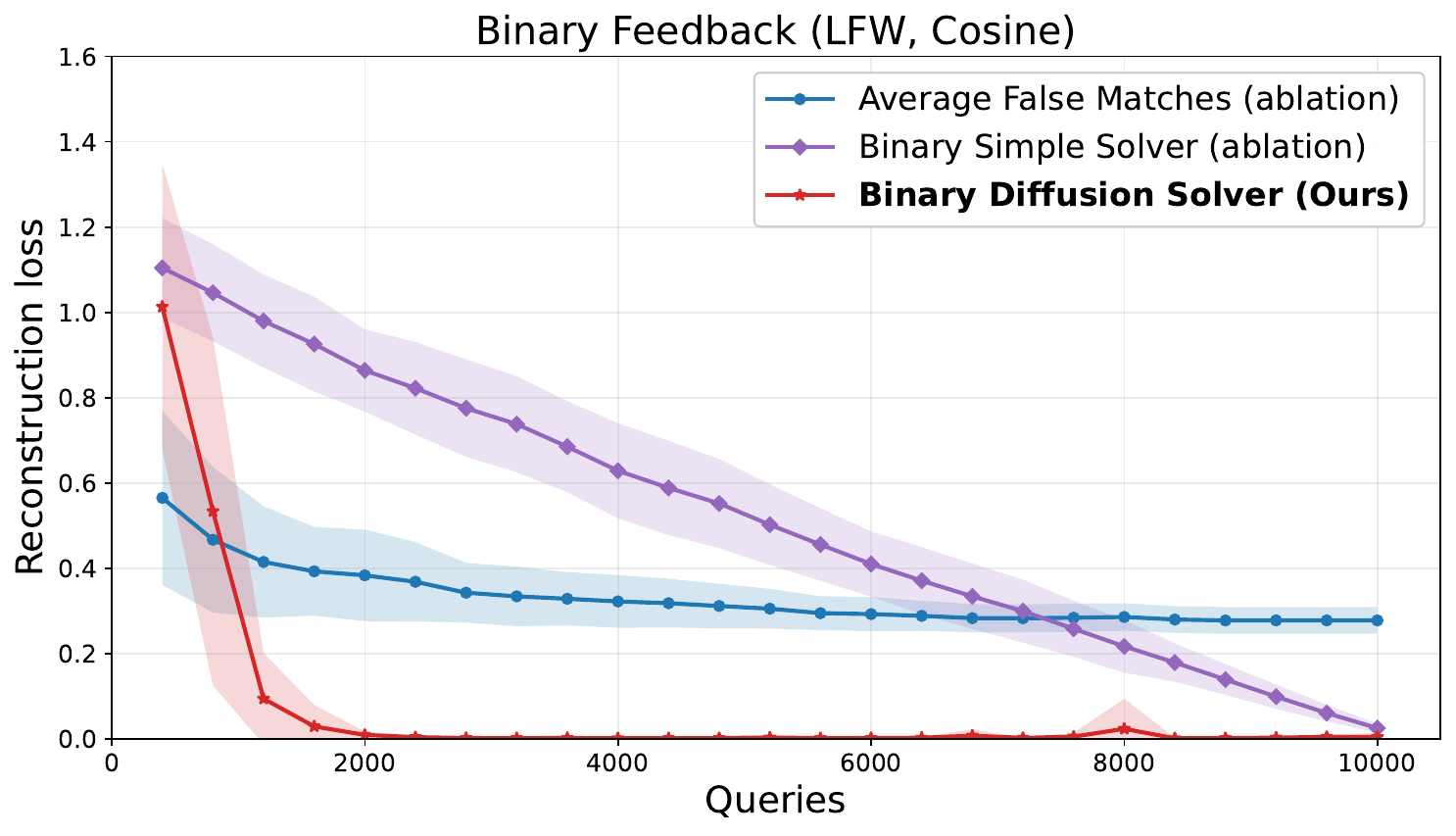}
    \hfill
    \includegraphics[width=0.48\textwidth]{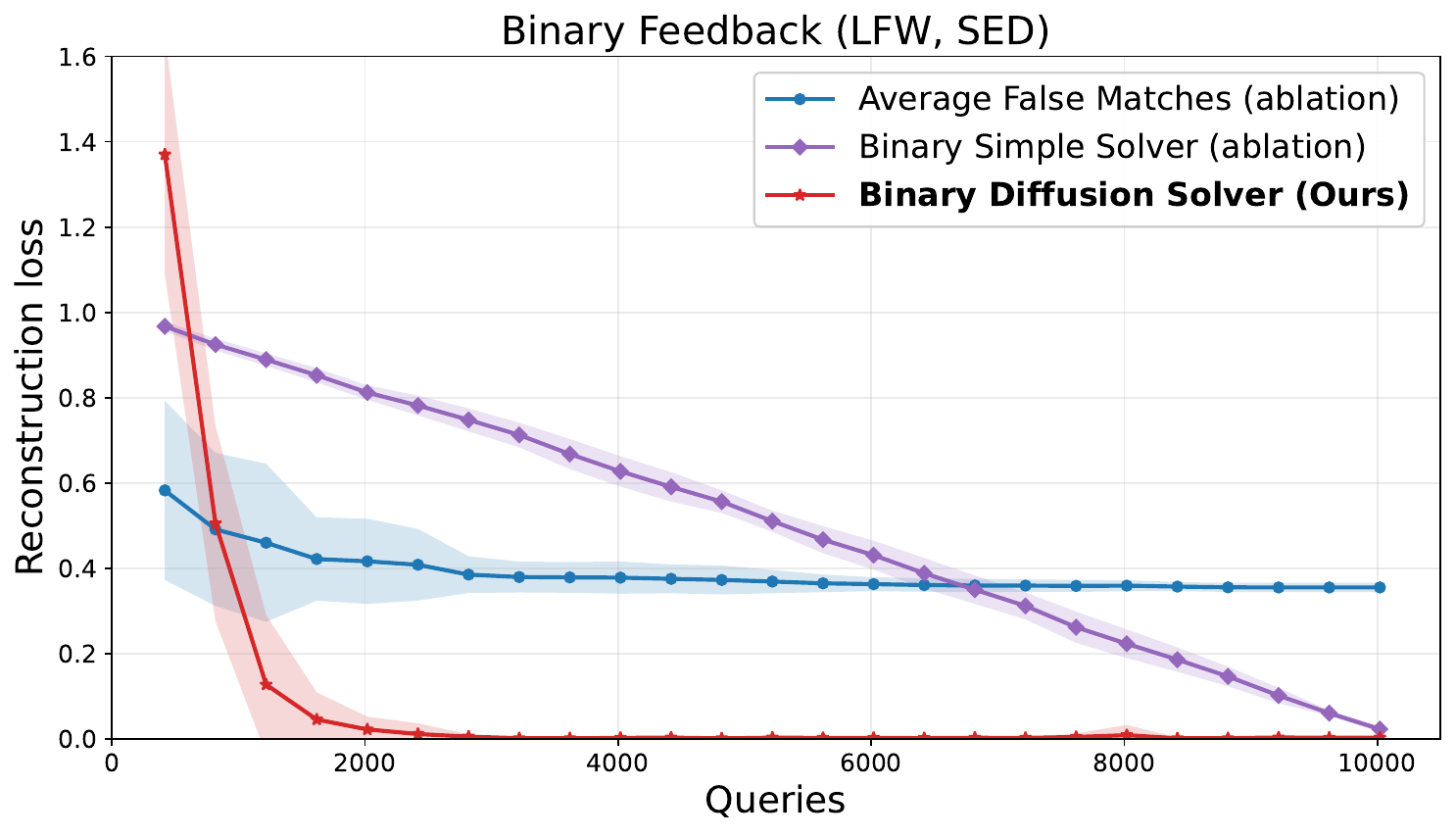}
    \caption{Additional benchmark and decision-boundary results for query-efficient binary reconstruction across CelebA and LFW under cosine and SED similarity measures. These results are consistent with the main-text results and further support our findings.}
    \label{fig:appendix_binary_reconstruction}
\end{figure*}

\subsubsection{Query-Efficient Reconstruction from Similarity Scores}
\label{app:scores_with}

Figure~\ref{fig:appendix_score_with_selection} presents additional results for query-efficient reconstruction from similarity scores with input selection. These results are consistent with the trends reported in the main text and further support our findings regarding the effectiveness of distribution-aware query selection.

\subsubsection{Query-Efficient Reconstruction from Similarity Scores Without Input Selection}
\label{app:scores_without}

Figure~\ref{fig:appendix_score_without_selection} presents additional results for reconstruction from similarity scores without input selection. These results are also consistent with the main-text results and further highlight the importance of input selection for improving query efficiency.

\subsubsection{Query-Efficient Binary Reconstruction}
\label{app:binary}

Figure~\ref{fig:appendix_binary_reconstruction} presents additional results for query-efficient binary reconstruction. As in the main text, the additional results consistently support the effectiveness of the proposed approach in reducing the number of required authentication queries.


\subsection{Extended Implementation Details}
\label{app:extended_implementation_details}

\subsubsection{Our Diffusion-Based Solver}
\label{app:diffusion_solver_details}

We incorporate a pretrained facial diffusion model as a generative prior for
template reconstruction. Given a linear constraint system $Af=b$ derived from
the authentication feedback, our solver guides diffusion sampling toward
images whose extracted biometric features satisfy these constraints.
Algorithm~\ref{alg:posterior_sampling} summarizes the procedure.

We implement the solver using the pretrained IDiff-Face latent diffusion
model~\cite{10378084}. Specifically, we use the official
25\% context-dropout model,
\texttt{unet-cond-ca-bs512-150K-cpd25}, trained for $150{,}000$ steps on
FFHQ. The model is used solely as a generative facial prior, without any
additional training or fine-tuning, and is invoked without identity
conditioning. At each denoising step, the predicted clean latent is decoded
using the pretrained IDiff-Face VQ decoder, and the corresponding biometric
template is extracted using the \texttt{InceptionResnetV1} FaceNet model
pretrained on VGGFace2.

Before diffusion sampling, we apply SVD row whitening to the linear
constraint system for both SED- and cosine-based reconstruction. Given
\begin{equation}
    A=U\Sigma V^\top,
\end{equation}
we retain the numerical-rank components and define
\begin{equation}
    \widetilde{A}=V_r^\top,
    \qquad
    \widetilde{b}=\Sigma_r^{-1}U_r^\top b.
\end{equation}
This transformation preserves the retained linear constraints while making
the corresponding measurement directions orthonormal. The resulting
feature-space conditioning objective is
\begin{equation}
    \mathcal{L}_i
    =
    \operatorname{MSE}(\widetilde{A}f_i,\widetilde{b})
    =
    \frac{1}{r}
    \left\|
        \widetilde{A}f_i-\widetilde{b}
    \right\|_2^2,
\end{equation}
where $r$ is the retained numerical rank. We use a relative singular-value
threshold of $10^{-10}$. Whitening is applied only to the system provided
to the diffusion solver; the algebraic reconstruction solver operates on
the original $(A,b)$ system.

Gradients of $\mathcal{L}_i$ are backpropagated through the biometric
feature extractor and the VQ decoder to the current noisy latent. We use
the raw FaceNet feature directly in the conditioning loss for both SED and
cosine constraints. Sampling is initialized from a Gaussian latent and
uses the deterministic DDPM posterior-mean update followed by feature-space
guidance. We use $T=1000$ denoising steps, a fixed guidance scale
$\lambda=1000$, switching timestep $\tau=300$, and
$\varepsilon=10^{-8}$. For $i\geq\tau$, the guidance magnitude is
normalized according to
\begin{equation}
    \zeta_i
    =
    \lambda
    \frac{\sigma_i}
         {\|\nabla_{x_i}\mathcal{L}_i\|_2+\varepsilon},
\end{equation}
whereas for the final $300$ denoising steps we use
$\zeta_i=\lambda$.

All experiments use a single generated sample and CUDA execution. The
solver returns the FaceNet embedding of the predicted clean image from the
final denoising iteration.

\subsubsection{Lagrange-Based Reconstruction Method}

We evaluate the Lagrange-based reconstruction baseline of~\cite{10449211} using the authors' publicly available implementation.\footnote{\url{https://github.com/aminabassit/tra-he-btps}.} For each query budget, we generate
independent random probes from a standard Gaussian distribution and normalize
each probe to a unit norm. The released scores are then converted into the linear
system
\begin{equation}
    A x \approx b.
\end{equation}
The reconstructed template is obtained by solving
\begin{equation}
    \hat{x}
    =
    \arg\min_x \|Ax-b\|_2^2
    \quad
    \text{s.t.}
    \quad
    \|x\|_2=1.
\end{equation}
Introducing a Lagrange multiplier $\lambda$ gives
\begin{equation}
    \hat{x}(\lambda)
    =
    \left(A^\top A+\lambda I\right)^{-1}A^\top b.
\end{equation}
We numerically select $\lambda$ such that
$\|\hat{x}(\lambda)\|_2^2=1$. The root-finding procedure is initialized with
$\lambda=0.1$, and the resulting linear system is solved using least squares.
The recovered vector is additionally normalized to unit norm. If the numerical
solution for $\lambda$ fails, we use the unconstrained least-squares solution
followed by unit normalization.

\subsubsection{GMM-Based Prior}

As an additional distribution-aware baseline, we estimate the feature-space
distribution using a Gaussian mixture model (GMM) trained on the
training-validation embeddings. We use one mixture component with a full
covariance matrix and a fixed random seed of $0$. Thus, in our implementation,
the GMM corresponds to a full-covariance Gaussian estimate of the target
feature distribution.

Given the score-derived system $Ax\approx b$, reconstruction is formulated as
maximum a posteriori estimation:
\begin{equation}
    \hat{x}
    =
    \arg\min_x
    \frac{1}{2\sigma^2}\|Ax-b\|_2^2
    -
    \log p_{\mathrm{GMM}}(x).
\end{equation}
We set $\sigma^2=1$ and initialize the optimization using the ordinary
least-squares solution
\begin{equation}
    x_{\mathrm{init}}
    =
    \arg\min_x\|Ax-b\|_2^2.
\end{equation}
The objective is optimized using L-BFGS-B for a maximum of $20$ iterations.
The GMM is trained once for each evaluated dataset and feature extractor and is
shared across all attacked target templates.

\subsubsection{Hill Climbing}

We implement hill climbing as an iterative score-guided search following prior
score-leakage attacks. The initial reconstruction
$x^{(0)}$ is sampled from a standard Gaussian distribution and normalized to
unit norm. At iteration $t$, we sample a random unit direction $u_t$ and
construct the candidate
\begin{equation}
    x'
    =
    \frac{x^{(t)}+\eta u_t}
    {\|x^{(t)}+\eta u_t\|_2}.
\end{equation}
The candidate is retained only when its released matching score improves over
the current reconstruction; otherwise, the previous reconstruction is kept.
For SED, an improvement corresponds to a lower distance, while for normalized
templates, minimizing SED is equivalent to maximizing cosine similarity. We use
a fixed step size of
\begin{equation}
    \eta=0.07
\end{equation}
throughout the attack, without step-size decay. Each candidate evaluation is
counted as one authentication query, and the reported values correspond to the
cumulative number of hill-climbing iterations.

\subsubsection{PCA-Based Input Selection}

For each target identity, we construct a breaking-set of 10,000 images that do
not belong to the target identity and extract their unit-normalized biometric
templates. We apply PCA to these templates and select the eigenvectors
associated with the largest eigenvalues according to a variance threshold
$\tau$. In our experiments, we set $\tau$ such that the first 40 principal
components are retained. These directions capture the greatest variation
in the biometric feature-space distribution and therefore provide
informative and non-redundant constraints. For query budgets larger than
40, the remaining queries are selected randomly from the breaking-set.
The released scores are then converted into linear constraints, and the
target template is reconstructed using least squares.

\end{document}